\documentclass[longauth]{aa} 

\usepackage{graphicx}
\usepackage{txfonts}
\usepackage{multirow}
\usepackage{booktabs}
\usepackage{xcolor}
\usepackage{gensymb}
\usepackage{textcomp}
\usepackage{natbib}
\usepackage{hyperref}

\begin{document}

   \title{The GAPS Programme at TNG XXI \thanks{
   Based on observations made with the Italian {\it Telescopio Nazionale Galileo} (TNG) operated by the {\it Fundaci\'on Galileo Galilei} (FGG) of the {\it Istituto Nazionale di Astrofisica} (INAF) at the {\it  Observatorio del Roque de los Muchachos} (La Palma, Canary Islands, Spain). Partly based on data obtained with the STELLA robotic telescopes in Tenerife, an AIP facility jointly operated by AIP and IAC.}} 
\subtitle{A GIARPS case-study of known young planetary candidates: confirmation of HD 285507 b and refutation of AD Leo b}

   \author{I. Carleo\inst{1,2},
          L. Malavolta\inst{3},
          A.~F.~Lanza\inst{3},
          M. Damasso\inst{4},
          S. Desidera\inst{2},
          F. Borsa\inst{5},
          M. Mallonn\inst{6},
          M. Pinamonti\inst{4},
          R. Gratton\inst{2},
          E. Alei\inst{2},
          S. Benatti\inst{7},
          L. Mancini\inst{8,9,4},
          J. Maldonado\inst{7},
          K. Biazzo\inst{3},
          M. Esposito\inst{10},
          G. Frustagli\inst{5,22},
          E. Gonz{\'a}lez-{\'A}lvarez\inst{11},
          G. Micela\inst{7},
          G. Scandariato\inst{3},
          A. Sozzetti\inst{4},
          L. Affer\inst{7},
          A. Bignamini\inst{12},
            A. S. Bonomo\inst{4},
            R. Claudi\inst{2},
            R. Cosentino\inst{13},
            E. Covino\inst{14},
            A. F. M. Fiorenzano\inst{13},
            P. Giacobbe\inst{4},
            A. Harutyunyan\inst{13},
            G. Leto\inst{3},
            A. Maggio\inst{7},
            E. Molinari\inst{15},
            V. Nascimbeni\inst{2},
            I. Pagano\inst{3},
            M. Pedani\inst{13},
            G. Piotto\inst{16},
            E. Poretti\inst{13,5},
            M. Rainer\inst{17},
            S. Redfield\inst{1},
            C. Baffa\inst{17},
            A. Baruffolo\inst{2},
            N. Buchschacher\inst{18},
            V. Billotti\inst{17},
            M. Cecconi\inst{13},
            G. Falcini\inst{17},
            D. Fantinel\inst{2},
            L. Fini\inst{17},
            A. Galli\inst{13},
            A. Ghedina\inst{13},
            F. Ghinassi\inst{13},
            E. Giani\inst{17},
            C. Gonzalez\inst{13},
            M. Gonzalez\inst{13},
            J. Guerra\inst{13},
            M. Hernandez Diaz\inst{13},
            N. Hernandez\inst{13},
            M. Iuzzolino\inst{19},
            M. Lodi\inst{13},
            E. Oliva\inst{17},
            L. Origlia\inst{20},
            H. Perez Ventura\inst{13},
            A. Puglisi\inst{17},
            C. Riverol\inst{13},
            L. Riverol\inst{13},
            J. San Juan\inst{13},
            N. Sanna\inst{17},
            S. Scuderi\inst{3},
            U. Seemann\inst{21},
            M. Sozzi\inst{17},
            A. Tozzi\inst{17}
          }

\institute{Astronomy Department and Van Vleck Observatory, Wesleyan University, Middletown, CT 06459, USA  \\ 
              \email{icarleo@wesleyan.edu}
\and INAF - Osservatorio Astronomico di Padova, Vicolo dell'Osservatorio 5, I-35122, Padova, Italy 
\and INAF - Osservatorio Astrofisico di Catania, Via S. Sofia 78, I-95123, Catania, Italy 
\and  INAF - Osservatorio Astrofisico di Torino, Via Osservatorio 20, I-10025, Pino Torinese, Italy 
\and INAF - Osservatorio Astronomico di Brera, Via E. Bianchi 46, 23807 Merate, Italy 
\and Leibniz-Institut f{\"u}r Astrophysik Potsdam (AIP), An der Sternwarte 16, D-14482 Potsdam, Germany 
\and INAF - Osservatorio Astronomico di Palermo, Piazza del Parlamento, 1, I-90134 Palermo, Italy 
\and Department of Physics, University of Rome ``Tor Vergata'', Via
della Ricerca Scientifica 1, I-00133, Rome, Italy 
\and Max Planck Institute for Astronomy, K\"{o}nigstuhl 17,
D-69117, Heidelberg, Germany 
\and Thuringer Landessternwarte Tautenburg, Sternwarte 5, 07778, Tautenburg, Germany 
\and Centro de Astrobiología (CSIC-INTA), Carretera de Ajalvir km 4, 28850 Torrejón de Ardoz, Madrid, Spain 
\and INAF - Osservatorio Astronomico di Trieste, Via Tiepolo 11, 34143 Trieste, Italy 
\and Fundaci{\'o}n Galileo Galilei-INAF, 
Rambla Jos{\'e} Ana Fernandez P{\'e}rez 7, 38712 Bre{\~n}a Baja, TF, Spain 
\and INAF - Osservatorio Astronomico di Capodimonte, Salita Moiariello 16, 80131 Napoli, Italy 
\and  INAF Osservatorio Astronomico di Cagliari \& REM, Via della Scienza, 5, I-09047 Selargius CA, Italy 
\and Dipartimento di Fisica e Astronomia Galileo Galilei, Universit{\'a} di Padova, Vicolo dellOsservatorio 3, I-35122, Padova, Italy 
\and INAF-Osservatorio Astrofisico di Arcetri
Largo Enrico Fermi 5, I-50125 Firenze, Italy 
\and Département d'Astronomie -- Université de Genève, Chemin des Maillettes, 51, CH-1290, Versoix, Switzerland 
\and Officina Stellare S.r.l., Via Della Tecnica, 87/89, I-36030 Sarcedo (VI) - Italy 
\and INAF – Osservatorio Astronomico di Bologna, Via Gobetti 93/3, I-40129, Bologna, Italy 
\and Institut f\"ur Astrophysik -- Georg-August-Universit\"at G\"ottingen, Friedrich-Hund-Platz 1, D-37077 G\"ottingen, Germany 
\and Dipartimento di Fisica G. Occhialini, Università degli Studi di Milano-Bicocca, Piazza della Scienza 3, 20126 Milano, Italy 
}

   \date{Received: 07 January 2020 ; accepted: 24 February 2020 }

 
  \abstract
   {The existence of hot Jupiters is still not well understood. Two main channels are thought to be responsible for their current location: a smooth planet migration through the proto-planetary disk or the circularization of an initial high eccentric orbit by tidal dissipation leading to a strong decrease of the semimajor axis. Different formation scenarios result in different observable effects, such as orbital parameters (obliquity/eccentricity), or frequency of planets at different stellar ages.}
   {In the context of the GAPS Young-Objects project, we are carrying out a radial velocity survey with the aim to search and characterize young hot-Jupiter planets. Our purpose is to put constraints on evolutionary models and establish statistical properties, such as the frequency of these planets from a homogeneous sample.
   }
   {Since young stars are in general magnetically very active, we performed multi-band (visible and near-infrared) spectroscopy with simultaneous GIANO-B + HARPS-N (GIARPS) observing mode at TNG. This helps to deal with stellar activity and distinguish the nature of radial velocity variations: stellar activity will introduce a wavelength-dependent radial velocity amplitude, whereas a Keplerian signal is achromatic. As a pilot study, we present here the cases of two already claimed hot Jupiters orbiting young stars: HD~285507\,b and AD~Leo\,b.}
   {Our analysis of simultaneous high-precision GIARPS spectroscopic data confirms the Keplerian nature of HD285507's radial velocities variation and refines the orbital parameters of the hot Jupiter, obtaining an eccentricity consistent with a circular orbit. On the other hand, our analysis does not confirm the signal previously attributed to a planet orbiting AD~Leo. This demonstrates the power of the multi-band spectroscopic technique when observing active stars.}
   {}

   \keywords{Instrumentation: spectrographs -- Planetary systems
                --
                Technique: spectroscopic -- Stars: activity -- Techniques: radial velocities -- Stars: individual: HD~285507, AD~Leo
               }
\titlerunning{}
\authorrunning{I. Carleo et al.}
   \maketitle
%

\section{Introduction}
\label{sec:intro}
The discovery of the first hot Jupiter (HJ), 51~Peg~b \citep{mayorandqueloz1995}, orbiting with a period of 4.23 days around its host star, challenged our assumptions about planetary formation, even inside our own Solar System. The subsequent discoveries of the Kepler mission (see e.g. \citealt{batalhaetal2014}), which revealed the diversity of planetary system architectures, clearly demonstrated that most of the main features of the Solar System are only one occurrence within a wide range of possibilities. This consideration strengthened the interest in the field of exoplanet discovery to obtain a large sample of well-characterized giant exoplanets. 

Different theories were developed in order to explain the origins and properties of HJs. These include in-situ formation \citep{batyginetal2016,boleyetal2016,maldonadoetal2018}, migration due to the Kozai mechanism \citep{Kozai}, disk migration \citep{linandpapaloizou1986}, planet-planet scattering \citep{chatterjeeetal2008} and secular chaos \citep{wuetal2011}. All of these scenarios are characterized by different timescales and are expected to produce observable effects that can be used to assess their relative effectiveness \citep{dawsonjohnson2018}. These include differences in orbital parameters (eccentricity and/or obliquity), age-dependent frequency of different systems architectures, and differences in atmospheric composition depending on formation site and migration history.

The search for planets around young stars across a wide range of stellar ages (from few Myrs to few hundreds of Myrs) provides an opportunity to study ongoing and recent planet formation. These systems can help evaluate the role played by various migration mechanisms, and the influence that formation site and orbit evolution have on the observed planetary diversity. 

To date, the sample of claimed young close-in exoplanets is limited. In addition to a handful of young directly imaged systems \citep{bowler2016}, there are six planets discovered with RV and/or transit techniques, with ages younger than $\sim$20 Myr: PTFO 8-8695 b (3 Myr, \citealt{vaneyken2012,yuetal2015}), V830 Tau b (2 Myr, \citealt{donatietal2016}), CI Tau b (2 Myr, \citealt{johnskrulletal2016}), K2-33 b (5-10 Myr, \citealt{davidetal2016,mannetal2016b}), TAP26 (10-15 Myr, \citealt{yuetal2017}), V1298 Tau b (23 Myr, \citealt{davidetal2019}).
As a matter of fact, past surveys avoided young stars in their samples due to their high activity level. The stellar activity, indeed, makes the detection of young HJs very difficult, since the planetary radial velocity (RV) signal is buried in the stellar noise. This issue has led to biased statistics or the retraction of previously claimed planets after subsequent investigations (e.g. \citealt{carleo2018}).

Although the number of HJs around young stars is small, it is still higher than expected based on statistics of older stars \citep{yuetal2017}. Extending the observed sample of exoplanets orbiting young stars is crucial to better understand planet formation. 

The direct imaging method can only detect relatively cold massive planets at wide separation from their host stars \citep{chauvin2018}, and therefore does not contribute to the HJs sample. High-resolution spectroscopy and RV monitoring together with the transit method 
are well suited to enlarge the HJ sample. The complementarity of these methods is useful for planet validation and provide diagnostics of stellar magnetic activity.
Simultaneous multi-band spectroscopy represents a powerful tool to overcome the contamination in the spectra caused by the stellar activity. RV jitter due to the activity is reduced in the NIR with respect to the VIS range, due to the smaller contrast between stellar spots and the rest of the stellar disk \citep{Prato, Mahmud, crockettetal2012}. Therefore, coupling near-infrared (NIR) and visible (VIS) observations can clearly identify the contribution attribute to activity. If the RV variation is solely due to Keplerian motion of a planet, the signal strength will be  wavelength-independent (e.g. \citealt{Gonzalez}, \citealt{benatti2018} and references therein).


In this paper, we present the first results of the GAPS (Global Architecture of Planetary Systems) Young Objects Project, based on the observations of stars with claimed planets. We report the RV analysis for two previously announced hot-Jupiter planets orbiting around HD~285507 \citep{quinnetal2014} and AD~Leo \citep{tuomietal2018}. These observations were performed in GIARPS mode \citep{claudietal2017}, which simultaneously uses HARPS-N \citep{cosentinoetal2012} in the VIS range (0.39 -- 0.68\,$\mu$m) and GIANO-B (the NIR high-resolution spectrograph at Telescopio Nazionale Galileo, TNG, \citealt{olivaetal2006}) in the NIR range (0.97 -- 2.45\,$\mu$m). Both spectrographs are mounted on the 3.6\,m Telescopio Nazionale Galileo in La Palma, Spain. Although these stars are not extremely young, this work demonstrates the power of the multi-band spectroscopy technique in distinguishing the activity signal from variation due to a planetary companions. \\

This paper is organized as follows: we describe the GAPS Young Objects Project and its objectives in Section \ref{sec:gapsyo}; 
we present the observations and analysis for HD~285507 in Section \ref{sec:yo13}, and for AD Leo in Section \ref{sec:yo17}. The interpretation of the data is presented in Section \ref{sec:disc}. We draw our conclusions in Section \ref{sec:concl}.


\section{GAPS Young Objects Project}
\label{sec:gapsyo}
The GAPS project \citep{covinoetal2013} started in 2012 as a radial velocity survey with HARPS-N (a cross-dispersed high-resolution echelle spectrograph, \citealt{cosentinoetal2012,cosentinoetal2014}) at the TNG. The aim was to determine the planet frequency around different types of host stars. A new phase of the project has been recently initiated by taking advantage of the unique capabilities of the GIARPS mode.
With this new facility, which obtains simultaneous multi-wavelength spectroscopy and can isolate the contribution by stellar activity, the GAPS project intends to explore planetary systems diversity with observations of planetary atmospheres \citep[][Pino et al. submitted]{borsa2019} and the detection of planets around young stars. With this purpose, the Young Objects Project searches for planets around young stars spanning from $\sim$ 2 Myr (e.g., Taurus star forming region) to 400-600 Myr ages (e.g., members of open clusters such as Coma Ber, and moving groups such as Ursa Major). Some known planet candidates 
are included in the sample for independent confirmation. 

A dense cadence of RV observations in individual observing season is mandatory for characterizing the effects of magnetic activity, using techniques such as Gaussian processes \citep{damasso2017,malavoltaetal2018} and kernel regression \citep{lanza2018}.
This is achieved thanks to the sharing of telescope time with similar observing programs (e.g. HARPS-N GTO).
In order to help the reconstruction of the behaviour of active regions of stellar surface, we also acquired photometric time series
of most of our targets using the STELLA (STELLar Activity) telescopes for northern targets (see Sec. \ref{sec:obsredyo17}) and REM (Rapid Eye Mount) at La Silla for equatorial targets \citep{carleo2018}.

With this sample, we expect to be able to confirm or disprove the significantly higher frequency of HJ at very young ages \citep{yuetal2017}. In addition, we hope to
distinguish among the timescales of the HJ migration, by comparing the frequencies and properties of planets at the age of few Myr 
to those at a few hundreds Myr, and to those orbiting older stars. 
Depending on the level of magnetic activity, we also expect to 
be sensitive to lower mass (Hot Neptunes, HNs) and longer period (warm Jupiters) planets around the intermediate-age stars (Malavolta et al. 2020, in prep.; Quinn et al. 2020 in prep.). These planets will enable us to investigate different regimes of planetary migration,
and to explore the possibility that a fraction of HNs evolved from partially evaporated HJs.
Our sensitivity to warm Jupiters nicely complements the detection parameter-space of NASA's TESS (Transiting Exoplanet Survey Satellite) satellite \citep{rickeretal2015}, which is limited by the 27 days duration of the observation for a large fraction of the sky. 
We have started to exploit the availability of short-period transit candidates around young stars provided by the TESS mission, with the validation of the 40 Myr old
planet around DS Tuc \citep{benatti2019,newtonetal2019}. This will ensure a better efficiency for the survey as opposed to a blind search. Furthermore, the knowledge of planetary radius from transits is critical for the physical characterization of the planet, providing additional diagnostics to disentangle the formation and migration path of close-in planets.




\section{HD~285507}\label{sec:yo13}
In this Section we present the RV measurements of the star HD~285507, a member of the Hyades open cluster (age=625 $\pm$ 50 Myr, \citealt{perrymanetal1997}, or a more recent estimation of 650 $\pm$ 70 Myr, \citealt{martinetal2018}).
It is a K4.5 star, with magnitudes $V$=10.5, $J$=8.4, $H$=7.8, $K$=7.7 \citep{cutri2003}, $v$\,sin\,$i$ = 3\,km\,s$^{-1}$ \citep{chaturvedi2016} and a rotational period of $11.98$ days \citep{delormeetal2011}. It is also a star with a relatively low activity level, according to the chromospheric index $log R'_{HK}$ = -4.54, calculated through the dedicated tool of the HARPS -- N DRS with the method provided in \citealt{lovis}, as implemented on YABI workflow (\citealt{yabi}).

\cite{quinnetal2014} claimed the presence of a slightly eccentric ($e$=0.086) hot Jupiter around this star. RVs from TRES spectra (Tillinghast Reflector Echelle Spectrograph at the Fred L. Whipple Observatory in Arizona, \citealt{furesz2008}) show a root mean square ($r.m.s.$) scatter of 89.5~m\, s$^{-1}$ and a period of 6.0881 days. Since no correlation between RVs and S-index was found, they attributed the origin of the RV variations with a semi-amplitude of 125.8~m\, s$^{-1}$ to a companion of minimum mass of 0.917\,$\rm M_{J}$. However, since the planet period is close to half the stellar rotational period, further confirmation of such a planet is necessary. Moreover, if confirmed, the moderate eccentricity could constrain which migration channel is responsible for the short period of this planet.

\subsection{Observations and data reduction}
We monitored this star in GIARPS mode between 2017 Oct 18 and 2018 Mar 31, in order to gather simultaneous multi-wavelength high resolution spectra. A summary of the available data from HARPS-N and GIANO-B and their properties are presented in Table \ref{tab:YO13summary}. The dataset is composed of 30 GIANO-B spectra and 18 HARPS-N spectra. The difference in the number of spectra between the two instruments is due to the different exposure times necessary to have a similar signal-to-noise ratio (SNR) from the two spectrographs, and reach the fixed detection integration time (DIT) for the NIR.

GIANO-B data were reduced with both the on-line Data Reduction Software (DRS) pipeline \citep{harutyunyanetal2018} and the off-line GOFIO pipeline \citep{rainer2018}.
\begin{table*}
\centering
\caption{\label{tab:YO13summary} Summary of the spectroscopic data of HD~285507. For each dataset we list the instrument used for the observations, the number of spectra, the median S/N, the RV nominal internal error ($\sigma_{RV}$) and the RV r.m.s. scatter.}
\begin{tabular}{lcccc}
\hline
\noalign{\smallskip}
Instrument    &  N$_{\rm spectra}$      & SNR     &   $\sigma_{RV}$     &    RV r.m.s.  \\ 
   &  &            &  (km\, s$^{-1}$)       &  (km\, s$^{-1}$)  \\ 
\hline  
\noalign{\smallskip}
GIANO-B &   30  &     39      &       0.061          &   0.119   \\
\noalign{\smallskip}
\hline
\noalign{\smallskip}
HARPS-N   &  18  &    42     &       0.003      &    0.105  \\
\noalign{\smallskip}
\hline
\end{tabular}
\end{table*}
The GIANO-B RVs (Table \ref{tab:YO13RVgiano}) were 
obtained by calculating the weighted average RV for each exposure and its corresponding error from the RVs of the individual orders \citep{carleoetal2016,carleo2018}. The HARPS-N RVs (Table \ref{tab:YO13RVharpsn}) were obtained with the HARPS-N DRS, using the K5 mask template, and also with the TERRA pipeline \citep{anglada2012}. Since TERRA RVs are consistent within the errors to the DRS values, hereinafter we only consider the DRS outputs for this target, having smaller errors. 
\\

\subsection{Data analysis and results}
\label{sec:analysisyo13}
\subsubsection{HARPS-N data}
\begin{figure}
$\begin{array}{cc}
\includegraphics[width=1.0\linewidth]{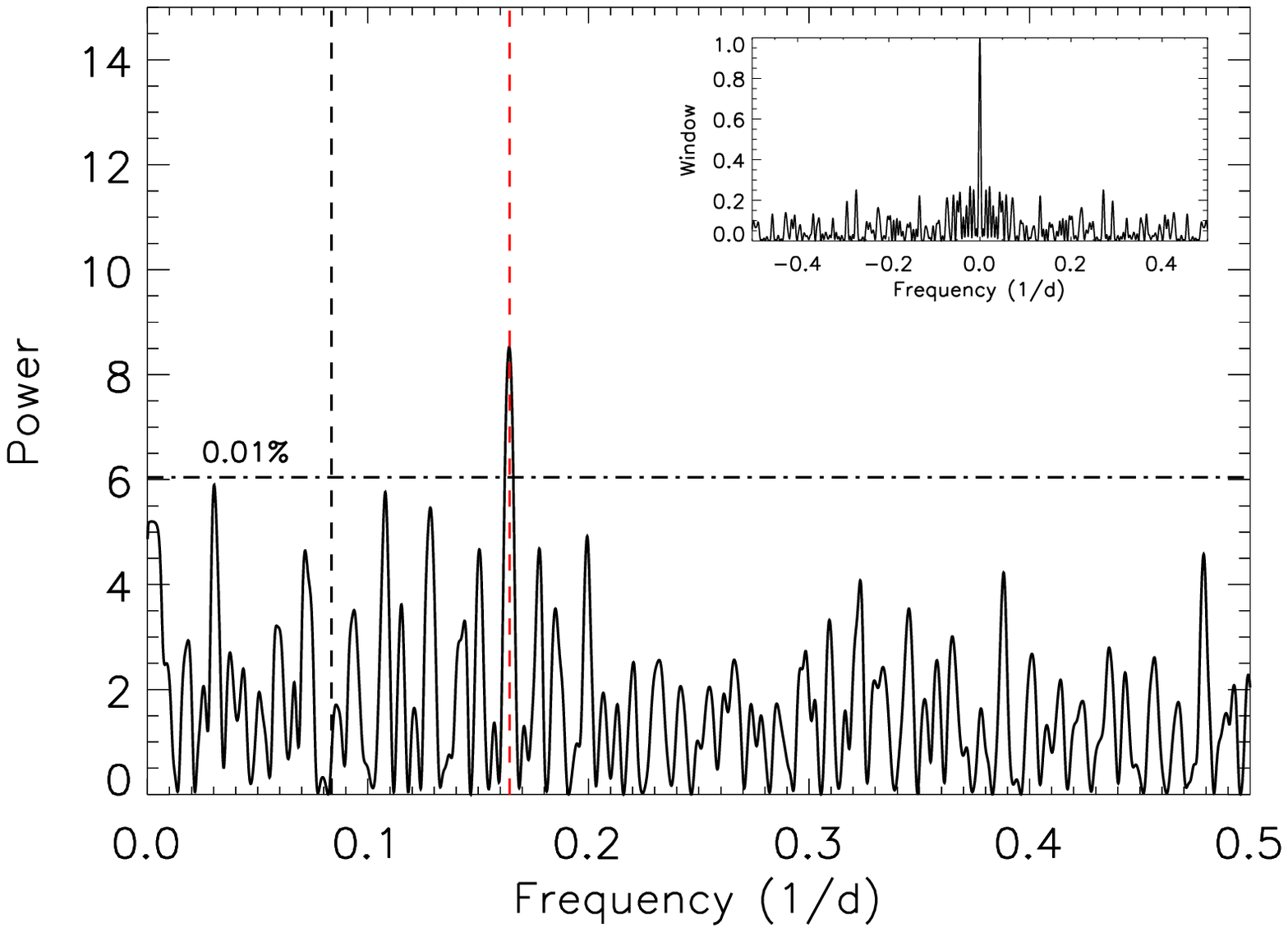}\\
\includegraphics[width=1.0\linewidth]{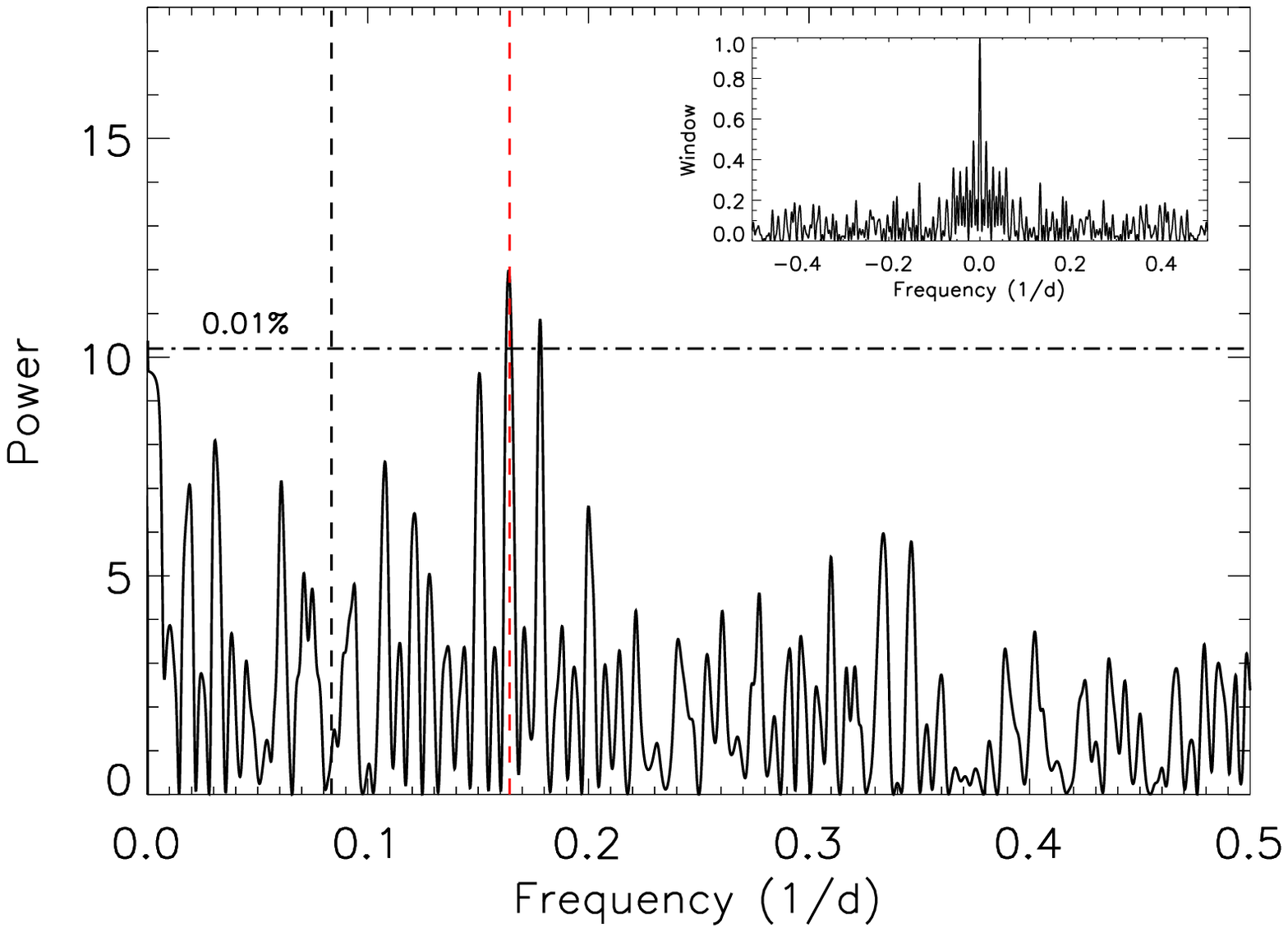} \\
\end{array} $
 \caption{\textit{Top}: GLS of HD~285507 for HARPS-N data. \textit{Bottom}: GLS of HD~285507 for GIANO-B data. The black vertical dashed line represents the rotational period at 11.98 days, while the red vertical dashed line marks the proposed planet period at 6.0881 days. The horizontal dot-dashed line indicates the power level corresponding to a FAP = 0.01\%.}
\label{fig:YO13_gls}
\end{figure}

We computed the Generalised Lomb-Scargle (GLS) periodogram \citep{zech09} for HARPS-N data (Fig. \ref{fig:YO13_gls}, upper panel), which exhibits a highly significant periodicity at 6.095 days, i.e. very close to the periodicity found by \cite{quinnetal2014}. The orbital fit is obtained from the VIS RVs, in a Bayesian framework using the package {\tt PyORBIT}\footnote{Available at \url{https://github.com/LucaMalavolta/PyORBIT}} (\citealt{malavoltaetal2016,malavoltaetal2018}). The parameter space was sampled using the affine invariant Markov chain Monte Carlo (MCMC) ensemble sampler {\tt emcee} \citep{foremanmackey2013}, with starting conditions obtained with the global optimization code {\tt PyDE}\footnote{Available at \url{https://github.com/hpparvi/PyDE}}. We run the sampler for 50000 steps with 44 walkers, for a total of 13200 posterior samplings per each parameter, after removing the first 20000 steps as burn-in and applying a thinning factor of 100.
The resulting RV semi-amplitude is $K$=137.9\,$\pm$\,4.4\,m\,s$^{-1}$, the period is $P$=6.0950$ _{-0.0046}^{+0.0043}$ days and the eccentricity $e$=0.043$ _{-0.028}^{+0.033}$, showing a good agreement with the result obtained by \cite{quinnetal2014}. 

The residuals from the VIS orbital fit have a scatter of 10~m\, s$^{-1}$ with a peak in the periodogram of 31 days that has a false alarm probability (FAP) of 23\,\%, i.e. not highly significant.
The FAP is estimated via a bootstrap method, generating 10,000 artificial RV curves obtained from the real data, keeping the epochs of observations fixed but making random permutations in the RV values.


\subsubsection{GIANO-B data}
The GLS and orbital fit analyses were also performed for GIANO-B data. The NIR periodogram (Fig. \ref{fig:YO13_gls}, bottom panel) shows a significant periodicity at 6.105 days. The small number of GIANO-B data and their larger errors in the RV measurements do not allow for a proper estimate of the eccentricity, hence we performed the NIR orbital fit assuming a circular orbit, motivated by the low eccentricity found in the HARPS-N data. Moreover, our main goal in this case was to verify the semi-amplitude of the planetary signal in the NIR, with respect to the VIS one. The resulting semi-amplitude is 143.4 $\pm$ 16.6~m\, s$^{-1}$. 
This value coincides with that obtained from HARPS-N spectra, Within the errors. After subtracting the NIR orbital fit, the r.m.s. of NIR residuals is 48.4~m\, s$^{-1}$, and the corresponding periodogram shows a non-significant peak at 28 days. 



Comparing the VIS and NIR datasets, HARPS-N and GIANO-B RVs show a similar RV  r.m.s., i.e., 105 and 119~m\, s$^{-1}$, respectively (Table \ref{tab:YO13summary}). Moreover, VIS and NIR semi-amplitudes are consistent within the error bars, confirming the Keplerian nature of the RV variation. 

\subsubsection{NIR-VIS joint fit}
We also performed a joint Keplerian fit of the TRES, HARPS-N and GIANO-B RVs.
Our model includes an eccentric orbit for the planetary signal, and independent offset and jitter terms for each dataset, for a total of 11 parameters. Following \cite{eastmanetal2013}, we sampled the orbital period ($P$) and the RV semi-amplitude ($K$) in logarithmic space, employing the $k = \sqrt{e} \cos {\omega_\star} $ and $h = \sqrt{e} \sin {\omega_\star}$ parametrization for the eccentricity ($e$) and the argument of periastron ($\omega_\star$). Instead of fitting for the central time of transit $T_C$ (not available in our case), we fitted the combination $\phi = \omega_\star + M_0$, where $M_0$ is the mean anomaly at an arbitrary  reference time T$_{\rm ref}$ = 2456399.0 \citep{ford2006}. A summary of orbital parameters obtained from the fitting of data taken with different instruments is reported in Table \ref{tab:YO13summaryfit}.

\begin{figure}
    \centering
    \includegraphics[width=1.0\linewidth]{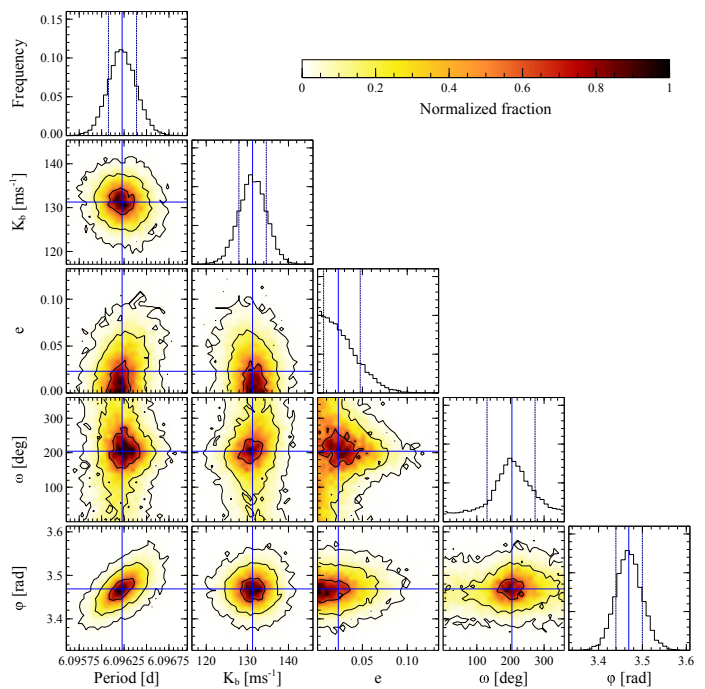}
    \caption{Correlations among the major orbital parameters obtained for HD~285507 from the MCMC fit.}
    \label{fig:YO13cornerplot}
    \vspace{-0.3cm}
\end{figure}

\begin{figure}
    \centering
    \includegraphics[width=0.90\linewidth]{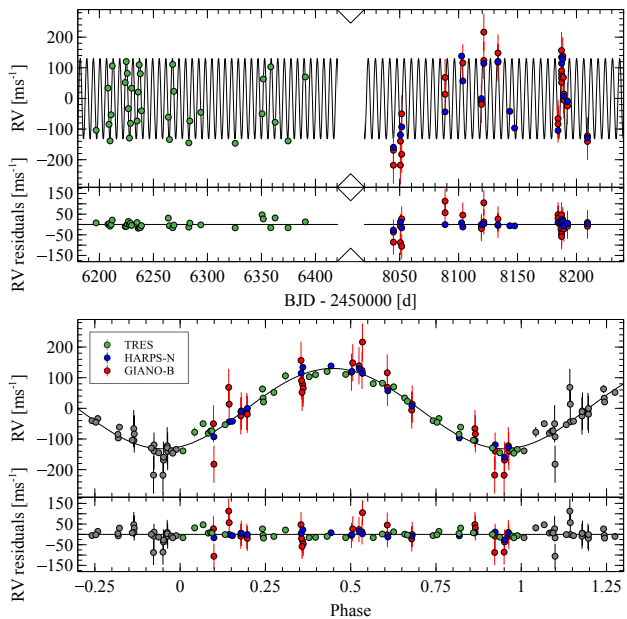}
    \caption{Orbital fit (black line) of HD~285507 at 6.0962 days obtained combining the visible data from HARPS-N (blue diamonds) and TRES (green dots), and GIANO-B  RVs (red dots) in the NIR.}
    \label{fig:YO13RVvisnir}
    \vspace{-0.3cm}
\end{figure}

\begin{figure}
    \centering
    \includegraphics[width=1.0\linewidth]{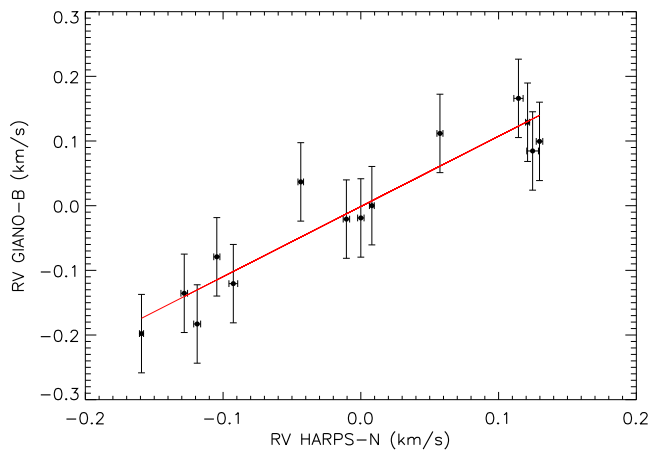}
    \caption{Correlation between simultaneous VIS and NIR RVs of HD~285507.}
    \label{fig:YO13correlation}
    \vspace{-0.3cm}
\end{figure}

The resulting orbital solution gives a RV semi-amplitude of $K$=131.3\,$\pm$\,3.5\,m\,s$^{-1}$, a period of $P$=6.0962\,$\pm$\,0.0002\,days and an eccentricity of $e$=$0.023 _{-0.015}^{+0.024}$, that is consistent within 1.5 $\sigma$ of a circular orbit. The corner plot in Fig. \ref{fig:YO13cornerplot} shows the correlations between the orbital parameters: no evident correlation is visible. The orbital model is represented in Fig. \ref{fig:YO13RVvisnir} as dashed line, while VIS and NIR RVs are overplotted. The median and 1\,$\sigma$ interval confidence of the parameters are reported in Table~\ref{tab:YO13summaryfit}.
The NIR and VIS RVs acquired simultaneously are almost all consistent with each other within 1$\sigma$.
This leads to the confirmation of the Keplerian nature of RV variations, also supported by a strong correlation between VIS and NIR RVs (Fig. \ref{fig:YO13correlation}), with a Spearman and Pearson correlation coefficients of 0.89 and 0.94, respectively. This correlation has been obtained considering only the common nights between the VIS and NIR datasets and averaging the VIS RVs in the same night. 

\subsubsection{Stellar parameters determination}
Finally, we determined the mass and radius of the star following the same approach described in \cite{2019MNRAS.484.3233B}. Briefly, with the \texttt{isochrone} package \citep{2015ascl.soft03010M} we performed an isochrone fit  using as priors the Gaia DR2 distance (d $= 45.03 \pm 0.25$ pc, \citealt{2018AJ....156...58B}), the photometry from the Two Micron All Sky Survey (2MASS; \citealt{cutri2003}; \citealt{2006AJ....131.1163S}), the Wide-field Infrared Survey Explorer (WISE;\citealt{2010AJ....140.1868W}), and Johnson $B$ and $V$ photometry, together with the photospheric parameters from \cite{quinnetal2014}. We constrained the age between 550 and 750 Myr, consistent with the value usually adopted for the Hyades \citep{perrymanetal1997}. For stellar models, we used both MESA Isochrones and Stellar Tracks (MIST; \citealt{2016ApJ...823..102C}; \citealt{2016ApJS..222....8D}; \citealt{2011ApJS..192....3P}) and the Dartmouth Stellar Evolution Database \citep{2008ApJS..178...89D}.  We obtained $M_\star = 0.771 \pm 0.006 M_\odot$ and $R_\star = 0.699 \pm 0.004 R_\odot$, which converts the RV semi-amplitude into a minimum mass of the planet of $M_p \sin{i} = 0.992 \pm 0.026 M_J$.


\begin{table*}[htbp]
\centering
\caption{\label{tab:YO13summaryfit} Summary of the orbital parameters of HD~285507 resulting from the fitting model for different instruments.}
\begin{tabular}{lcccc}
\hline
\noalign{\smallskip}
Instrument    &  P     & $K$ & $e$ & $\omega$  \\
             &  (d)      &  (m\, s$^{-1}$)   &             &    $\rm deg$    \\
\hline  
\noalign{\smallskip}
GIANO-B &   6.1037 $_{-0.0135}^{+0.0131}$  &  143.4 $\pm$ 16.6  & Circular  & -\\
\noalign{\smallskip}
\hline
\noalign{\smallskip}
HARPS-N &    6.0950$ _{-0.0046}^{+0.0043}$  &  137.9 $\pm$ 4.4 & 0.043$ _{-0.028}^{+0.033}$ & 297 $\pm$ 57\\
\noalign{\smallskip}
\hline
\noalign{\smallskip}
HARPS-N + GIANO-B + TRES &   6.0962 $\pm$ 0.0002  &  131.3 $\pm$ 3.5  & 0.023$ _{-0.015}^{+0.024}$ & 206 $\pm$ 75\\
\noalign{\smallskip}
\hline
\noalign{\smallskip}
TRES \citep{quinnetal2014} &   6.0881 $_{-0.0018}^{+0.0019}$  &  125.8 $\pm$ 2.3  & 0.086 $_{-0.019}^{+0.018}$ & 182 $\pm$ 11\\
\noalign{\smallskip}
\hline
\end{tabular}
\end{table*}

\section{AD Leo}\label{sec:yo17}
In this Section we present the RV measurements of AD Leo (Gl 388, M4.5Ve, $V$=9.5, $H$=4.8, \citealt{Reiners2013}). AD Leo is a very close (distance of 4.9 pc, \citealt{pettersen1981}) active star, with an estimated age between 25 and 300 Myr \citep{Shkolnik2009}, and $v$\,sin\,$i$ = 3~km\, s$^{-1}$ \citep{reinersetal2007}. AD Leo has an inclination of the rotation axis of $\sim$20$\degree$ according to \cite{morinetal2008} and $\sim$15$\degree$ according to \cite{tuomietal2018}, so it is seen nearly pole-on. 

Given its brightness and proximity, AD Leo is a very well studied star. \cite{hawleypettersen1991} observed AD Leo in the 1200-8000 \AA ~wavelength range, collecting photometric data with a standard two-star photometer attached to the 0.9 m telescope at McDonald Observatory, and spectroscopic data with the Electronic Spectrograph No. 2 at the 2.1 m Struve telescope at the same site. They also employed ultraviolet spectra obtained with the International Ultraviolet Explorer (IUE) satellite \citep{boggessetal1978}. This joint analysis reported a giant flare which lasted for more than 4 hours. 

\cite{HuntWalker2012}, using MOST (Microvariability and Oscillations of Stars satellite) observations, captured 19 flares in 5.8 days and found a sinusoidal modulation in the light curve with a period of 2.23 days, attributing it to the rotation of a stellar spot. This period is consistent with the 2.24 d rotational period found by \cite{morinetal2008} through the Zeeman Doppler Imaging (ZDI) analysis. A further confirmation of this rotational period came from the HARPS RV variations found by \cite{Bonfils}. Their periodogram analysis showed a peak at 2.22 days, and the strong correlation between RVs and bisectors demonstrated that stellar activity was the source of the RV variation.

Very similar results were found by \cite{Reiners2013}, studying the Zeeman effect on RV measurements, interpreting the 2.22704 d periodicity as originating from the corotation of star spots on the stellar surface, since the line asymmetries were correlated with the RV modulation. Very recently, \cite{lavailetal2018} reported a sudden change in the surface magnetic field of AD Leo, through the study of magnetic maps using ZDI from ESPaDOnS spectropolarimetric data. The significant change in the shape of the circular polarisation profiles was attributed to a decrease of the total magnetic field of about 20\% between 2012 and 2016.

Finally, \cite{tuomietal2018} presented a photometric and spectroscopic analysis of AD Leo, proposing the presence of a planet orbiting the star in a spin–orbit resonance in order to explain the RV variations from HARPS and HIRES spectra \citep{Vogt1994}. They found a photometric period of 2.22791$_{-0.00055} ^{+0.00066}$ days from ASAS photometry, and a spectroscopic period of 2.22579$_{-0.00023} ^{+0.00014}$ days. Although the photometric and spectroscopic periods are very close and a strong correlation between HARPS RVs and bisector values was found, they performed tests on colour and time invariance of the RVs. First they divided the 72 HARPS orders in six sets of 12 orders each and calculated the weighted mean velocities, finding the signal to be independent of the selected wavelength range. Then they subdivided HARPS spectra in three different temporal subsets, treating them as independent datasets over a period of 60 days and found a stable periodic variability, consistent with a time-invariant signal. On the basis of these results, they proposed that the RV variations are due to Keplerian motion (with a semi-amplitude of 19.1~m\, s$^{-1}$) rather than to the effects of stellar activity. \\


\subsection{Observations and data reduction}\label{sec:obsredyo17}
\indent We performed GIARPS observations over three months (April-June) in 2018 in order to clarify the nature of the RV variations in AD Leo, coupling VIS and NIR spectroscopy. The GIARPS dataset is composed of 42 HARPS-N spectra and 25 GIANO-B spectra. An additional set of HARPS-N observations has been acquired between November 2018 and January 2019, collecting 21 more spectra. 

For a consistent comparison between our data and that presented in the discovery paper, HARPS-N RVs were extracted with the TERRA pipeline, and are listed in Table \ref{tab:YO17RVharpsn}. Henceforth, the first and second HARPS-N datasets will be named HT1 and HT2, respectively. We also extracted the RVs with the HARPS-N DRS for comparison (HD1 and HD2).

GIANO-B data were reduced with both the on-line DRS pipeline and the off-line GOFIO pipeline and divided in two different datasets (hereafter called G1 and G2). Technical operations on the instrument, including a whole system re-alignment, were performed between the acquisition of the two datasets and introduced drifts in our data. The GIANO-B RVs (Table \ref{tab:YO17RVgiano}) were obtained with the method described in \cite{carleoetal2016}.
A summary of the available data from HARPS-N and GIANO-B and their properties are presented in Table \ref{tab:YO17summary}.

During the same period as the GIARPS observations, we obtained 34 nights of photometric observations of AD Leo with the robotic 1.2m STELLA telescope \citep{strassmeier04} and its wide-field imager WiFSIP \citep{weber12}. Each observing block consisted of four exposures in Johnson $V$ band with an exposure time of 6s, and four exposures in Cousins $I$ band with 3s exposure time. The photometric data for $V$ and $I$ bands are listed in Tables \ref{tab:YO17stellaV} and \ref{tab:YO17stellaI}, respectively.

We performed the data reduction following the steps detailed in \cite{mallonn18}. In brief, a bias and flatfield correction was applied by the STELLA data reduction pipeline, and aperture photometry was executed with \texttt{SExtractor}\footnote{https://sextractor.readthedocs.io/en/latest/Introduction.html}. 
Different selections of nearby comparison stars were tested. The 
photometric signal of the target was found to be insensitive with respect to the choice of those stars, and the final comparison star was taken according to the minimization of point-to-point dispersion. We rejected data points that deviated significantly from the average flux level, Point Spread Function (PSF) shape and width, and level of background flux. In the end, we were left with 116 individual measurements in the $V$ and $I$ filters.

\begin{table*}[htbp]
\centering
\caption{\label{tab:YO17summary} Summary of the spectroscopic data of AD Leo. For each dataset we list the instrument used for the observations, the number of spectra, the median SNR, the RV nominal internal error ($\sigma_{RV}$), and the RV r.m.s. scatter. We report the HARPS-N results from the TERRA pipeline. }
\begin{tabular}{lcccc}
\hline
\noalign{\smallskip}
Dataset    &  N$_{\rm spectra}$      & SNR     &   $\sigma_{RV}$     &    RV r.m.s.   \\
   &  &            &  (km\, s$^{-1}$)       &  (km\, s$^{-1}$) \\
\hline  
\noalign{\smallskip}
GIANO-B (G1) &   12  &     180      &       0.019          &   0.019   \\
\hline
\noalign{\smallskip}
GIANO-B (G2) &   13  &     186      &       0.017          &   0.019   \\
\hline
\noalign{\smallskip}
HARPS-N  (HD1) &  42  &    74     &       0.001      &    0.030  \\
\hline
\noalign{\smallskip}
HARPS-N  (HD2) &  21  &    78     &       0.001      &    0.016  \\
\hline
\noalign{\smallskip}
HARPS-N  (HT1) &  42  &    74     &       0.001      &    0.022 \\
\hline
\noalign{\smallskip}
HARPS-N  (HT2) &  21  &    78     &       0.001      &    0.011 \\
\noalign{\smallskip}
\hline
\end{tabular}
\end{table*}


\subsection{Data analysis and results}
As mentioned in Sec. \ref{sec:obsredyo17}, our data consist of GIARPS spectra, plus an additional HARPS-N dataset, that are analysed separately and presented in the next sections. 

\subsubsection{GIARPS dataset}\label{subsubsec:giarpsyo17}

\subsubsection*{\textbf{HARPS-N data}}

\begin{figure}
    \centering
    \includegraphics[width=1.0\linewidth]{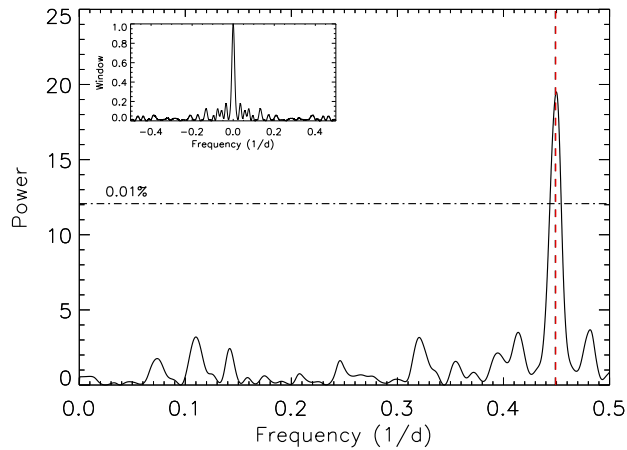}
    \caption{GLS of AD Leo for HARPS-N data (HT1). The black vertical dashed line is the rotational period at 2.22791 days, while the red vertical dashed line is the proposed planet period at 2.22579 days (both periods by \citealt{tuomietal2018}). The two lines overlap each other.}
    \label{fig:YO17_gls_harpsn_TERRA}
    \vspace{-0.3cm}
\end{figure}

The HARPS-N periodogram of the HT1 dataset (Fig. \ref{fig:YO17_gls_harpsn_TERRA}) exhibits a highly significant periodicity at 2.2231 days. The fitting model (using the package {\tt PyORBIT})
shows a period of 2.2244 $\pm$ 0.0010 days, an RV semi-amplitude of 33.0 $\pm$ 0.7~m\, s$^{-1}$, and an eccentricity of 0.176$ _{-0.019}^{+0.018}$. After subtracting the orbital fit, the residuals  r.m.s is 2.6~m\, s$^{-1}$, with no significant periodicity.

\begin{figure}
    \centering
    \includegraphics[width=1.0\linewidth,trim=0cm 0cm 0cm 0.75cm,clip]{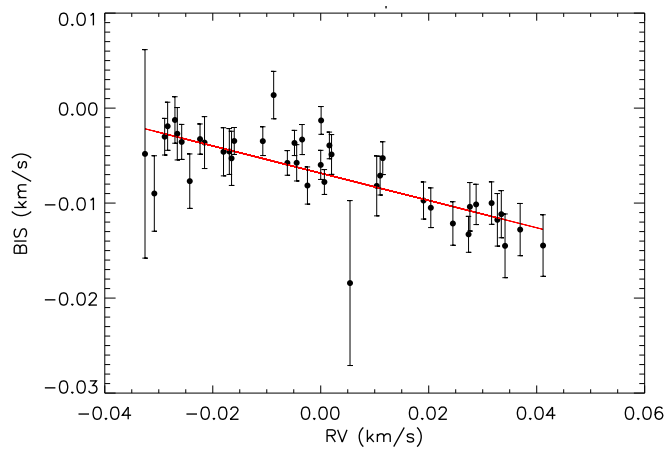}
    \caption{Correlation between VIS-TERRA RVs and BIS of AD Leo.}
    \label{fig:YO17correlation}
    \vspace{-0.3cm}
\end{figure}

\begin{figure}
    \centering
    \includegraphics[width=1.0\linewidth]{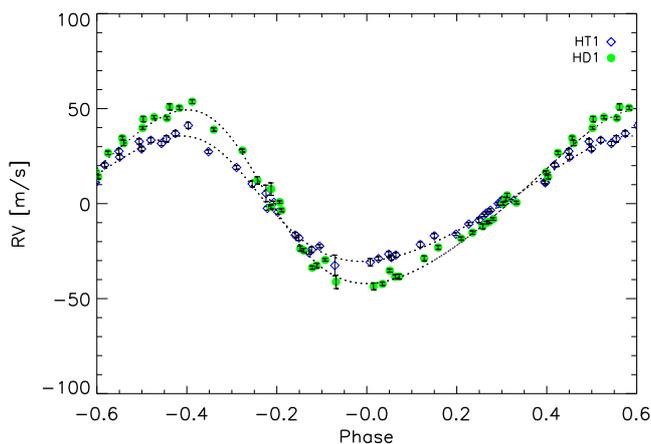}
    \caption{Comparison between DRS (green dots, HD1) and TERRA (blue diamonds, HT1) RVs with the corresponding Keplerian fits (dotted lines) phased at 2.2244 days.}
    \label{fig:YO17drsVSterra}
    \vspace{-0.3cm}
\end{figure}

We also investigated the correlation between the TERRA RVs and the BIS \citep{Queloz2001} obtained with the HARPS-N DRS, in order to understand the variability of the line profile due to stellar activity. Fig. \ref{fig:YO17correlation} shows a strong correlation (Spearman correlation coefficient of -0.76, Pearson correlation coefficient of -0.74), indicating that the RV variations could be due to stellar activity rather than a Keplerian signal. 
This result is also supported by the fact that, when comparing the RVs obtained by the DRS and the TERRA pipelines, they show different amplitudes. Fig. \ref{fig:YO17drsVSterra} shows the RVs provided by the two different RV calculation methods with the corresponding Keplerian fits: the modulation amplitude is lower for TERRA RVs. 

\subsubsection*{\textbf{GIANO-B data}}

The GLS and orbital fit analysis was also performed for GIANO-B data. The NIR periodogram 
shows a low significant periodicity at 11.5228 days,
and our fit of GIANO RVs does not reveal the presence of any significant periodicity. We then performed another fit by using a Gaussian prior of 2.225 $\pm$ 0.020 days on the period and limiting the eccentricity to 0.3, based on the results from the HARPS-N fit, in order to assess the detection limit of the HARPS-N signal in the GIANO dataset. We found a 3$\sigma$ upper limit on the semi-amplitude of the signal of 23~m\, s$^{-1}$ (99.7th percentile of the distribution), clearly incompatible with the 33.0 $\pm$ 0.7~m\, s$^{-1}$ measured with HARPS-N.
A summary of Keplerian fit parameters obtained from data of different instruments and pipelines is reported in Table \ref{tab:YO17summaryfit}.



\begin{figure}
    \centering
    \includegraphics[width=1.0\linewidth]{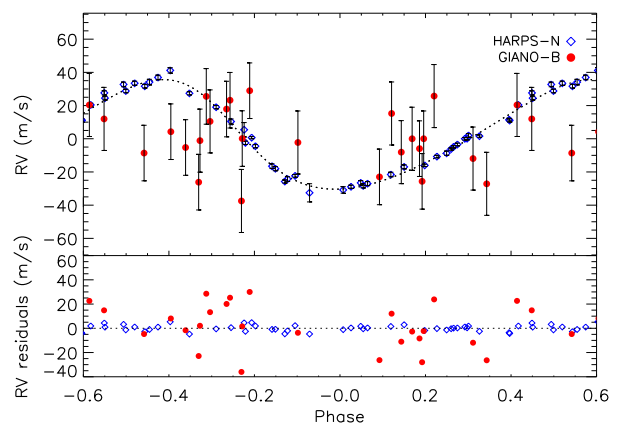}
    \caption{Keplerian fit (black dotted line) at 2.2244 days obtained with the visible data (HARPS-N, blue diamonds), with GIANO-B  RVs (red dots) overplotted.}
    \label{fig:YO17RV}
    \vspace{-0.3cm}
\end{figure}

Fig. \ref{fig:YO17RV} shows the VIS orbital fit together with the phase-folded VIS and NIR RVs. The NIR RVs do not lie on the VIS orbital curve. 
This strongly argues against a Keplerian explanation for the RV variations. This conclusion is also supported by the strong correlation between VIS RVs and BIS (Fig. \ref{fig:YO17correlation}). The r.m.s. of the RVs corrected for this trend is 14~m\, s$^{-1}$ as compared to the original r.m.s. of 22~m\, s$^{-1}$.  \\

\subsubsection{Additional HARPS-N dataset and confirmation of the activity-induced RV variation}
\begin{figure}
    \centering
    \includegraphics[width=1.0\linewidth]{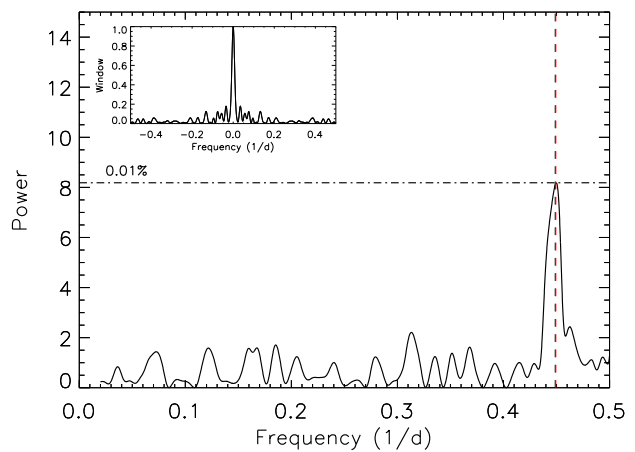}
    \caption{GLS of AD Leo for the second HARPS-N dataset (HT2). The black vertical dashed line indicates the maximum period of the NIR periodogram at 2.2238 days.}
    \label{fig:YO17_gls_harpsn_TERRA_data2}
    \vspace{-0.3cm}
\end{figure}

\begin{figure}
    \centering
    \includegraphics[width=1.0\linewidth]{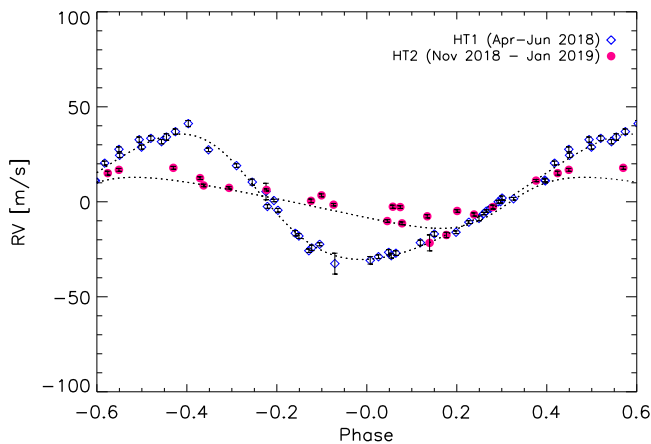}
    \caption{Comparison between the VIS RVs obtained from TERRA pipeline in the two different observing seasons (HT1, blue diamonds, and HT2 in pink dots).}
    \label{fig:YO17_dataset2}
    \vspace{-0.3cm}
\end{figure}

The result from the multi-wavelength observations, described in Sec. \ref{subsubsec:giarpsyo17}, is further supported by the last additional set of HARPS-N observations (HT2), that shows a significantly lower amplitude in comparison with the previous VIS dataset.
The periodogram of the HT2 dataset (Fig. \ref{fig:YO17_gls_harpsn_TERRA_data2}) exhibits a significant periodicity at 2.2238 days, while the fitting model has a period of 2.2225\,$\pm$\,0.0044 days and an RV semi-amplitude of 13.4\,$\pm$\,1.4~m\, s$^{-1}$. After subtracting the model fit, the RV residuals have an r.m.s. of 4.1~m\, s$^{-1}$ with a insignificant peak in the corresponding periodogram at 39 days.
In Fig. \ref{fig:YO17_dataset2} a comparison between HT1 and HT2 is shown. The change in amplitude and phase between the two periods is clear and demonstrates the non-Keplerian origin of the RV variation.

\subsubsection{STELLA dataset}
\begin{figure}
    \centering
    \includegraphics[width=1.0\linewidth, trim=0cm 10cm 0cm 3cm, clip]{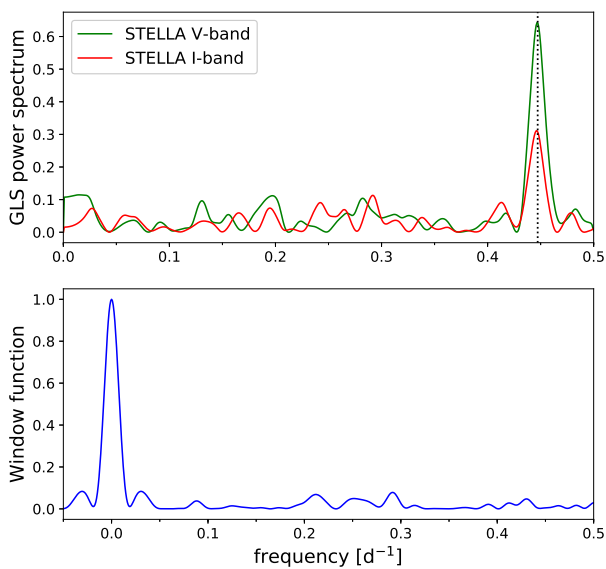}
    \caption{\textit{Upper plot.} GLS periodograms of the STELLA photometric data. The stellar rotation period $P=2.237\pm0.035$ days is clearly detected in both bands (and marked by a vertical dotted line), together with the 1-day alias. \textit{Bottom plot.} Window function of the STELLA photometric data.}
    \label{fig:YO17phot1}
    \vspace{-0.3cm}
\end{figure}

We investigated the GLS of the STELLA light curves (see Fig. \ref{fig:YO17phot1}), acquired quasi-simultaneously to the HT1 dataset. We found that the period with the highest power occurs at $P_{\rm rot,\:phot}=2.237\pm0.035$ days, calculated fitting a Gaussian function to the main peak of the periodogram\footnote{using \texttt{LMFIT}, a non-linear least-squares minimization and curve-fitting for Python.}, and adopting the width of the Gaussian profile as the uncertainty. The periodic modulation is more strongly detected in $V$-band photometry than in $I$-band, indicating that it is mainly due to dark spots on the stellar photosphere.

\begin{figure}
    \centering
    \includegraphics[width=1.0\linewidth, trim=0cm 2cm 0cm 2cm, clip]{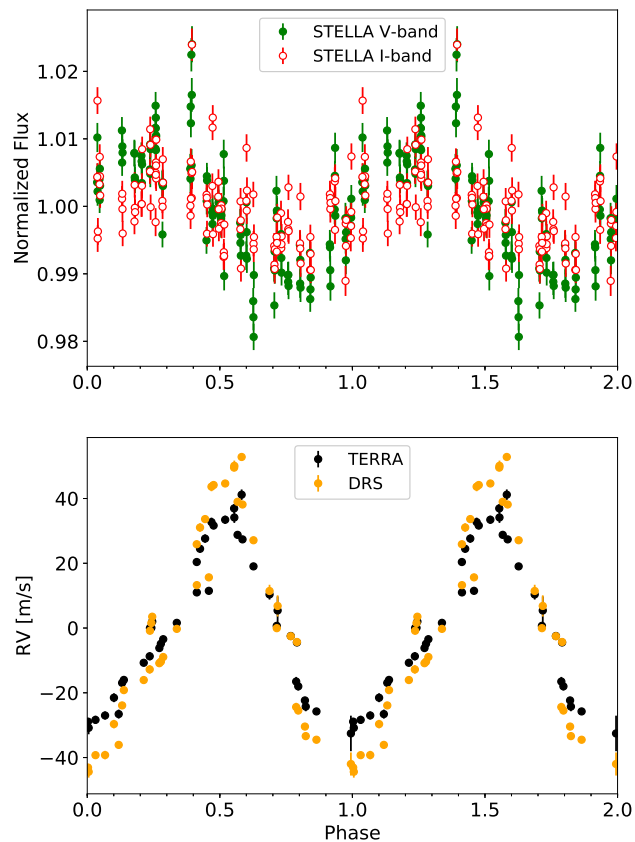}
    \caption{\textit{Upper plot.} STELLA photometric data phase folded to the stellar rotation period $P=2.237$ days found with GLS. \textit{Bottom plot.} Radial velocities of AD Leo, extracted with the TERRA and DRS pipelines, phase folded using the same period and phase of photometry.}
    \label{fig:YO17phot2}
    \vspace{-0.3cm}
\end{figure}

Fig. \ref{fig:YO17phot2} shows a comparison between the nearly-simultaneous photometric and RV curves both phase folded at the rotation period $P_{\rm rot,\:phot}$ and using the same epoch as the zero-point in phase. A clear shift of $\sim$0.25 in phase ($\sim$0.6 days) is evident between the photometric and RV curves.

\vspace{1cm}

\begin{table*}[htbp]
\centering
\caption{\label{tab:YO17summaryfit} Summary of the Keplerian parameters of AD Leo resulting from the fitting model.}
\begin{tabular}{lcccc}
\hline
\noalign{\smallskip}
Instrument/Pipeline    &  Period     & $K$  & $e$ & $\omega$ \\
                       &  (days)         &  (m\, s$^{-1}$)    &             &    ($deg$)      \\
\hline  
\noalign{\smallskip}
GIANO-B (G1+G2) &   2.2246 $\pm$ 0.0219 &   \textless\, 23   & - & -\\
\hline
\noalign{\smallskip}
HARPS-N (HT1) &   2.2244 $\pm$ 0.0010  &  33.0 $\pm$ 0.7  & 0.176$ _{-0.019}^{+0.018}$ & 63 $\pm$ 6 \\
\noalign{\smallskip}
\hline
\noalign{\smallskip}
HARPS-N (HT2) &   2.2225 $\pm$ 0.0044  &  13.4 $\pm$ 1.4  & 0.330 $\pm$ 0.089 & 262 $\pm$ 22 \\
\noalign{\smallskip}
\hline
\end{tabular}
\end{table*}

\section{Discussion}\label{sec:disc}

\subsection{HD~285507}
The analysis of RV measurements of the star HD~285507, a member of the Hyades open cluster, confirms the presence of the hot Jupiter planet announced by \cite{quinnetal2014}, since HARPS-N and GIANO-B RVs show similar semi-amplitudes within the error bars. 
 
Among the hot Jupiters with measurements of the orbital eccentricity, HD 285507 b stands out as one of the few planets with a well-determined age, making it a reference for studying the mechanisms of formation of hot Jupiters \citep{dawsonjohnson2018}. One may be tempted to take the circular orbit of HD 285507b as evidence of formation through type II migration in a circumstellar disc (e.g., \citealt{linandpapaloizou1986}). However, we caution that the large uncertainties in the efficiency of tidal dissipation inside hot Jupiters and their host stars prevent us from drawing conclusions in any single case (cf. \citealt{bonomoetal2017}). We aim at extending the sample of useful systems through our GAPS project.


\subsection{AD Leo}
We employed multi-wavelength spectroscopy to investigate the presence of a hot Jupiter around the very active star AD Leo, as proposed by \cite{tuomietal2018}. Our GIARPS data show that the modulation of the VIS RVs is not reproduced by the simultaneous NIR RVs, strongly rejecting the Keplerian nature of the VIS RV variations.

In addition, we detected significant changes in the amplitude and eccentricity of the Keplerian best fit to the optical RV modulation, that are not compatible with the presence of a planet. Stellar magnetic activity is therefore the best candidate to account for the optical RV variations. We calculated the minimum-mass detection threshold of the combined HARPS-N and GIANO-B RV datasets, following the Bayesian approach from \citet{tuomi14}. This Bayesian technique was applied adapting the publicly available \texttt{emcee} algorithm by \citet{foremanmackey2013}. Considering planets with periods between 1 and 100 d, we are sensitive to planets of minimum masses $M_{p}sini > 15.5$ $\rm M_{\oplus}$, for $1 < P < 10$ d, and $M_{p}sini > 29.5$ $\rm M_{\oplus}$ for $10 < P < 100$ d. It is worth noting that the short-period detection threshold is well below the mass of the candidate planet $M_{p} = 75.4 \pm 14.7$ $\rm M_{\oplus}$ proposed by \citet{tuomietal2018}.

\cite{tuomietal2018} investigated the available photometric time series finding a modulation at the rotation period with an amplitude of 9.3 mmag in the $V$ passband data before about JD~2455000, while no rotational modulation exceeding 6.7 mmag was observed after that date. With an amplitude of the optical photometric modulation of 10 mmag and a $vsini$ of 3 km\, s$^{-1}$, using Eq. (1) in \cite{desortetal2007}, we estimate an RV modulation with an amplitude half of that observed with HARPS-N. 
In the NIR the effects of activity on the RV are reduced, but not completely absent. This is because the Zeeman effect may produce RV variations that can counterbalance the reduced contrast of starspots as shown by Reiners et al. (2013). In the case of AD Leo, this introduces a further difficulty in explaining the optical RV variation  with a starspot. The model in Fig. 6 of \cite{Reiners2013} predicts an increase of the RV perturbation due to a magnetic spot with a temperature deficit of 700 K in the NIR, except in the $K$ band, as a consequence of the Zeeman effect. Indeed, very cool spots with a temperature deficit larger than 1200 K show a rather constant RV perturbation with increasing wavelength because of the reduced emission in the spotted photosphere that compensates for the Zeeman effect. In conclusion, in the case of a magnetic spot, we should expect a detectable modulation also in the GIANO-B RV measurements, which is not the case. Furthermore, the analysis by \cite{tuomietal2018} shows that the light curve of AD Leo is variable on timescales as short as a few days, suggesting that the fairly stable RV modulation has a different origin than the photometric modulation produced by starspots. 

The slow change of the optical RV modulation is better explained by considering the properties of the magnetic field of AD~Leo, which is dominated by a kG dipole aligned with the rotation axis (\citealt{morinetal2008}; \citealt{lavailetal2018}). Spectropolarimetric observations from 2006 to 2016 have been used to build ZDI maps of the photospheric field that demonstrate a remarkable stability in its long-term geometry. The strong dipole, being axisymmetric and aligned with the rotation axis, cannot induce a rotational modulation of the RV of AD~Leo. On the other hand, extended patches of meridional or azimuthal fields with an intensity of a few hundred Gauss can modify the photospheric convective flows and induce apparent shifts in the spectral lines akin to the quenching of line-shifts observed in solar faculae (cf. \citealt{meunieretal2010}). These patches appear in the ZDI maps of \cite{morinetal2008} and \cite{lavailetal2018} and are not associated with a reduced brightness of the photosphere because their fields are not strong enough to produce a local cooling as in starspots. 

Adopting a line shift of 200~m\, s$^{-1}$ in those magnetized patches, similar to the amplitude observed in solar faculae, we can account for the $\sim$65m\, s$^{-1}$ amplitude of the optical RV modulation with a filling factor of about 35 percent of the stellar disc. We can assume an average intensity of 200~G for the diffuse field in the patches, in agreement with the ZDI maps \citep{lavailetal2018}. The corresponding RV modulation produced by the Zeeman effect can be estimated by means of Eq.~(3) of \cite{Reiners2013}, and amounts to 1.1 m\,s$^{-1}$ at 500 nm and 12 m\,s$^{-1}$ at 1670 nm, that is the average wavelength of GIANO-B observations. Given the typical error of $\pm 20$~m\, s$^{-1}$ for our NIR RV measurements, the modulation is undetected in our NIR RV time series. A stronger field in the patches would imply a stronger quenching of the line shifts, thus requiring a smaller filling factor to reproduce the optical RV amplitude. This will compensate for the stronger Zeeman effect in the patches, thus maintaining the NIR RV modulation below our detection threshold.

\begin{figure}
    \centering
    \includegraphics[width=1.0\linewidth]{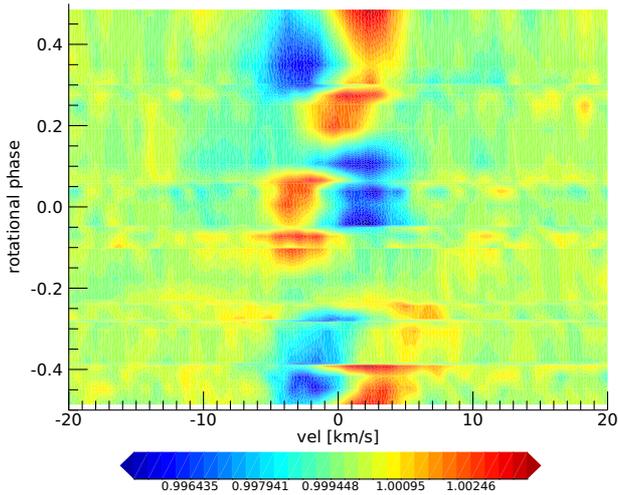}
    \caption{Contour map of the residuals of the CCF of AD Leo versus radial velocity and rotational phase.}
    \label{fig:contmapyo17}
    \vspace{-0.3cm}
\end{figure}

The contour map of the residuals of the cross-correlation function (CCF)\footnote{The CCF is provided by YABI by comparing the spectra with a line mask model.} of AD Leo versus radial velocity and phase is shown in Fig. \ref{fig:contmapyo17}. It shows bumps, that is positive deviations with respect to the mean CCF profile (in red), as well as dips, i.e. negative deviations (in blue). They can be associated with spots that are cooler and hotter than the unperturbed photosphere, respectively. Their excursion in radial velocity covers the full $\pm 3$ km\, s$^{-1}$ range of the $v \sin i$, indicating that those regions reach the edge of the stellar disk and are not confined to high latitudes close to the pole. The amplitude of the RV variation produced by the associated perturbation of the intensity can be estimated as $\Delta V_{\rm R} \simeq 2 \times v\sin i \times \Delta I \times f$, where $\Delta I \sim 0.008$ is the difference in intensity and $f < 1$ the filling factor. Therefore, they can account for a variation of $42 \times f$ m\, s$^{-1}$ that is insufficient to explain the observed amplitude of $\sim 90$ m\, s$^{-1}$ (cf. Fig. \ref{fig:YO17drsVSterra}). This supports the interpretation of the RV variation as mainly due to the quenching of convective blueshifts in magnetized regions of the photosphere rather than to the flux perturbations in dark spots or bright faculae.

 \section{Conclusions}
 \label{sec:concl}
In this paper, we have presented the first results of the new  RV survey performed in the framework of the GAPS project, focused on the detection of young planets.
The observations will be used to estimate orbital parameters and masses, in order to analyse the observable tracers of dynamical evolution and understand the origin of the observed diversity of planetary systems. 

We confirmed the presence of a hot Jupiter around the Hyades star HD 285507, showing that the amplitude of the RV variation is identical in VIS and NIR RVs. This represents an important verification since the planetary orbital period is close to the first harmonic of the rotational period of the star.
With our new analysis we also refined the value of the orbital eccentricity, consistent with zero which, in principle, suggests that a type II migration mechanism could be responsible for the current orbit of HD~285507 b.

We also demonstrated the chromaticity of the RV variations of the nearby, active M dwarf AD Leo, ascribing the observed modulation to stellar activity rather than a planetary companion.
 
 These first results demonstrate the great potential of multi-band high resolution spectroscopy when observing active stars, also attested by the study of BD+20 1790, based on GIARPS commissioning data complemented by GIANO-A and HARPS-N archive datasets \citep{carleo2018}. In the NIR regime stellar activity is reduced, but not completely absent, and an RV signal from the NIR band alone might lead to erroneous conclusions. For this reason, for very active stars it is critical to couple different wavelength ranges, such as VIS and NIR, in order to compare the corresponding results, and disentangle the nature of the RV variations.
 Our study also shows the relevance of independent observations and data processing for the
 confirmation of planets around young, active stars. While the field of planet detection using the RV technique is rather mature when considering chromospherically quiet stars, resulting in a limited number of controversial cases, the detection of planets
 at young ages is more challenging, making independent studies particularly welcome.

\begin{acknowledgements}
The authors acknowledge support by INAF/WOW and INAF/FRONTIERA through the "Progetti Premiali" funding scheme of the Italian Ministry of Education, University, and Research.  P.Gi. acknowledges financial support from the Italian Space Agency (ASI) under contract 2014-025-R.1.2015 to INAF. M.Pi. gratefully acknowledges the support from the European Union Seventh Framework Programme (FP7/2007-2013) under Grant Agreement No. 313014 (ETAEARTH).
\end{acknowledgements}



\newpage
\onecolumn
\begin{appendix}
\section{Tables}

\begin{longtable}{llcccc}
\caption{\label{tab:YO13RVgiano} Time series of HD~285507 from GIANO-B data. For each observation we list radial velocities (RV) and the bisector span (BIS) with the corresponding uncertainties.}\\
\hline
\noalign{\smallskip}
JD-2450000  &      RV         &    $\sigma_{\rm RV}$     &   BIS     & $\sigma_{\rm BIS}$  \\
    & (km\, s$^{-1}$) &    (km\, s$^{-1}$)   &  (km\, s$^{-1}$)   &  (km\, s$^{-1}$)  \\
\hline
\noalign{\smallskip}
8044.67522 &  -0.119         &    0.061   &    0.048         &      0.076\\
8044.68365 &  -0.070         &    0.061  &    0.089     &          0.078\\
8050.60310 &  -0.119         &    0.061   &   0.053     &          0.112\\
8050.61132 &  -0.040         &    0.061   &        0.003     &          0.123\\
8051.67397 &  0.049         &    0.061    &         0.030     &          0.086 \\
8051.68230 & -0.083         &     0.061  &        -0.014      &         0.212\\
8088.52388 &  0.168         &     0.061  &        0.017      &         0.076\\
8088.53215 &  0.113         &     0.061  &        0.011      &         0.072\\
8103.55010 &  0.215         &     0.061  &        0.017      &         0.100\\
8119.32899 &  0.089         &     0.061  &        0.026     &          0.073\\
8119.33730 &  0.080         &     0.061  &        0.057      &         0.071\\
8121.38722 &  0.223         &     0.061  &        0.033      &         0.080 \\
8121.39611 &  0.316        &     0.061   &        -0.006      &         0.108 \\
8133.39123 &  0.217        &     0.061  &        0.003     &          0.066 \\
8133.41360 &  0.248         &     0.061  &        0.067     &          0.067\\
8184.36706 &  0.033         &     0.061  &        0.021     &          0.088 \\
8184.37531 &  0.016       &     0.061    &        0.046      &         0.091\\
8187.35584 &  0.258       &     0.061   &       0.054     &          0.133\\
8187.36514 &  0.190       &     0.061   &       0.074      &         0.125\\
8187.37424 &  0.151       &     0.061   &       0.169      &         0.126\\
8187.38543 &  0.178       &     0.061   &       -0.035      &         0.136\\
8187.39495 &  0.165       &     0.061    &       0.173     &          0.154\\
8188.39191 &  0.238       &     0.061   &       0.079      &         0.076\\
8188.40016 &  0.168       &     0.061   &       0.066      &         0.081\\
8189.33426 &  0.093       &     0.061   &       -0.025     &          0.082\\
8189.34257 &  0.113       &     0.061   &       0.017      &         0.084\\
8192.37649 &  0.075       &     0.061   &        0.102      &         0.085\\
8192.38473 &  0.091       &     0.061  &       0.093       &        0.092\\
8209.34898 &  -0.023       &     0.061  &       0.015     &          0.161\\
8209.35722 &  -0.041       &     0.061  &        0.031     &          0.138\\
\hline
\noalign{\smallskip}
\end{longtable}
\vspace{1cm}

\begin{longtable}{lcccc}
\caption{\label{tab:YO13RVharpsn} Time series of HD~285507 from HARPS-N data. For each observation we list radial velocities (RV) and the bisector span (BIS) with their related uncertainties.}\\
\hline
\noalign{\smallskip}
 JD-2450000  &      RV         &    $\sigma_{\rm RV}$      & BIS    & $\sigma_{\rm BIS}$   \\
             & (km\, s$^{-1}$) &    (km\, s$^{-1}$)   &   (km\, s$^{-1}$)   &  (km\, s$^{-1}$) \\
\hline             
\noalign{\smallskip}
8044.68501  &  37.913 & 0.002  &    0.04893  &  0.003 \\
8050.61281  &   37.953 & 0.003   &    0.02831  &  0.005 \\
8051.68237  &  37.979 & 0.003  &    0.02613  &  0.005\\
8088.53361  & 38.028  & 0.002  &   0.03475  &  0.004 \\
8102.54544  &   38.210 & 0.003 &    0.04093  &  0.006 \\
8103.56040  &   38.129 & 0.002  &    0.05498  &  0.004 \\
8119.33795  &   38.072 & 0.002  &   0.01909  &  0.005 \\
8121.39496  &   38.186 & 0.003  &   0.05618  &  0.007 \\
8133.39928  &   38.193 & 0.001 &   0.05813  &  0.003 \\
8143.45555  &   38.030 & 0.003   &  0.01783  &  0.007 \\
8147.51170  &   37.976 & 0.005 &   0.04695  &  0.009 \\
8184.37034  &   37.967 & 0.002   &   0.03755  &  0.004 \\
8187.35909  &   38.187 & 0.004   &   0.02248  &  0.009 \\
8187.38850  &   38.206 & 0.005 &   0.06235  &  0.009 \\
8188.39561  &   38.202 & 0.002  &   0.03938  &  0.005 \\
8189.33710  &   38.080 & 0.002   &   0.05129  &  0.004 \\
8192.37866  &   38.061 & 0.002   &    0.01740  &  0.005 \\
8209.35035  &   37.944 & 0.002   &    0.04097  &  0.005 \\
\hline
\noalign{\smallskip}
\end{longtable}

\begin{longtable}{llrcrc}
\caption{\label{tab:YO17RVharpsn} Time series of AD Leo from HARPS-N data. For each observation we list radial velocities (RV) from TERRA pipeline and the bisector span (BIS) with their related uncertainties calculated through the YABI workflow.}\\
\hline
\noalign{\smallskip}

  \multirow{2}{*}{Dataset} & \multirow{2}{*}{JD-2450000}   &      RV         &    $\sigma_{\rm RV}$  &     BIS   & $\sigma_{\rm BIS}$\\ 
      &       &   (km\, s$^{-1}$)   &  (km\, s$^{-1}$) & (km\, s$^{-1}$)   &  (km\, s$^{-1}$)  \\
\hline             
\noalign{\smallskip}
 (HT1, Apr - Jun 2018) &   8213.45462  & -0.027 & 0.001 &  -0.003 & 0.002\\
  &  8214.45708  & 0.029 & 0.001 & -0.010 & 0.001 \\
  &  8215.41277 & -0.033 & 0.005 & -0.005 & 0.007 \\
  &  8216.45089  & 0.011 & 0.001 & -0.005 & 0.002 \\
  &  8236.47339  & 0.011 & 0.001 & -0.007 & 0.002\\
  &  8238.39345  & -0.006 & 0.001 & -0.006 & 0.001\\
  &  8238.40626  & -0.005 & 0.001 & -0.004 & 0.001\\
  &  8238.42611  & -0.003 & 0.001 & -0.003 & 0.001\\
  &  8238.54064  & 0.002 & 0.001 & -0.004 & 0.001\\
  &  8242.40770  & -0.027 & 0.001 & -0.001 & 0.003\\
  &  8243.42065  & 0.033 & 0.001 & -0.011 & 0.002\\
  &  8243.49769  & 0.034 & 0.002 & -0.015 & 0.003\\
  &  8244.50529  & -0.031 & 0.002 & -0.009 & 0.004\\
  &  8245.41667  & 0.020 & 0.001 & -0.010 & 0.002\\
  &  8245.48734  & 0.028 & 0.001 & -0.010 & 0.002\\
  &  8246.42625  & -0.026 & 0.001 & -0.004 & 0.002\\
  &  8248.44569  & -0.002 & 0.001 & -0.008 & 0.002\\
  &  8248.49771  & -0.004 & 0.001 & -0.006 & 0.002\\
  &  8249.44169  & -0.011 & 0.001 & -0.003 & 0.002\\
  &  8249.49209  & -0.009 & 0.001 & 0.001 & 0.003\\
  &  8251.42651  & -0.022 & 0.001 & -0.004 & 0.003\\
  &  8251.49558  & -0.017 & 0.001 & -0.005 & 0.002\\
  &  8252.43995  & 0.037 & 0.001 & -0.013 & 0.002\\
  &  8252.50337  & 0.041 & 0.002 & -0.014 & 0.002\\
  &  8253.44379  & -0.029 & 0.001 & -0.003 & 0.002\\
  &  8253.50703  & -0.028 & 0.001 & -0.002 & 0.002\\
  &  8254.38854  & 0.024 & 0.001 & -0.012 & 0.003\\
  &  8254.48513  & 0.033 & 0.001 & -0.012 & 0.003\\
  &  8266.37927  & -0.017 & 0.001 & -0.005 & 0.003\\
  &  8266.39689  & -0.018 & 0.001 & -0.005 & 0.002\\
  &  8266.46007  & -0.024 & 0.001 & -0.008 & 0.002\\
  &  8267.38229  & 0.000 & 0.001 & -0.006 & 0.002\\
  &  8267.39404  & 0.000 & 0.001 & -0.001 & 0.002\\
  &  8267.40140  & 0.002 & 0.001 & -0.005 & 0.002\\
  &  8268.39068  & 0.010 & 0.002 & -0.008 & 0.004\\
  &  8268.45878  & 0.005 & 0.004 & -0.018 & 0.007\\
  &  8269.39814  & -0.016 & 0.001 & -0.003 & 0.002\\
  &  8270.39871  & 0.027 & 0.001 & -0.013 & 0.002\\
  &  8272.39093  & 0.032 & 0.001 & -0.010 & 0.002\\
  &  8275.39709  & -0.022 & 0.001 & -0.003 & 0.001\\
  &  8277.39561  & 0.001 & 0.001 & -0.008 & 0.001\\
  &  8279.43523  & 0.019 & 0.001 & -0.010 & 0.002\\
\noalign{\smallskip}
\hline
\noalign{\smallskip}
 (HT2, Nov 2018 - Jan 2019) & 8438.75284  &  -0.009 &  0.003  & 0.015  &  0.001 \\
  &  8448.75849    &  -0.002     &     0.001 &   -0.007 &   0.002 \\
  &  8449.75883     & 0.011     &     0.001 &  -0.007 &   0.002\\
  &  8451.76844     & -0.002    &       0.001 &   -0.004 &   0.003\\
  &  8453.67788     & -0.022     &      0.004 &   -0.002 &   0.002\\
  &  8453.76103     & -0.017     &      0.001 &   -0.004 &   0.001\\
  &  8474.63715     & 0.018     &     0.001 &   -0.003 &   0.002\\
  &  8481.76498     & 0.006    &      0.001 &    -0.002  &   0.002\\
  &  8482.79033     & -0.007    &      0.001 &    -0.004  &   0.002 \\
  &  8483.80276     & 0.007     &     0.001 &    -0.005 &   0.002  \\
  &  8484.65702     & -0.011     &     0.001 &    -0.001 &   0.002\\
  &  8486.80670     & -0.010    &      0.001 &    -0.004  &   0.002 \\
  &  8487.70356     &  0.017    &      0.001 &    -0.010 &   0.002\\
  &  8488.70446     & 0.003     &     0.001 &    -0.009 &   0.002 \\
  &  8492.56687    & 0.009     &     0.001 &     -0.010  &   0.002 \\
  &  8495.76092    & -0.003    &      0.001 &     -0.003  &   0.002\\
  &  8502.56190    & -0.008    &      0.001 &      0.001  &   0.002\\
  &  8503.66212     & 0.012    &      0.001 &     -0.008  &   0.002\\
  &  8504.61442      & -0.003   &       0.001 &    -0.002  & 0.002\\
  &  8508.65540     & 0.001    &      0.001 &    -0.008 &   0.003\\
  &  8511.59974    & -0.005     &     0.001 &     -0.005 &  0.002\\
\noalign{\smallskip}
\hline
\noalign{\smallskip}
\end{longtable}
\vspace{1cm}
\begin{longtable}{llcccc}
\caption{\label{tab:YO17RVgiano} Time series of AD Leo from GIANO-B data. For each observation we list radial velocities (RV) and the bisector span (BIS) with the corresponding uncertainties.}\\
\hline
\noalign{\smallskip}
Dataset & JD-2450000  &      RV         &    $\sigma_{\rm RV}$     &   BIS     & $\sigma_{\rm BIS}$  \\
  &  & (km\, s$^{-1}$) &    (km\, s$^{-1}$)   &  (km\, s$^{-1}$)   &  (km\, s$^{-1}$)  \\
\hline
\noalign{\smallskip}
GIANO-B & 8238.41969 &  0.117         &    0.019      &    0.003         &      0.015\\
(G1) & 8238.53404 &  0.143         &    0.019     &    -0.028     &          0.014\\
 & 8243.41453 &  0.137         &    0.019                &   -0.059     &          0.015\\
 & 8243.49110 &  0.129         &    0.019                  &        0.001     &   0.019\\
 & 8244.49939 &  0.115         &    0.019                  &         -0.032     & 0.021 \\
 & 8245.41037 & 0.105         &     0.019                 &        -0.009      &   0.017\\
 & 8245.48146 &  0.090         &     0.019                 &        -0.007      &  0.018\\
 & 8246.43035 &  0.080         &     0.019                 &        -0.034      &  0.014\\
 & 8248.43926 &  0.116         &     0.019                 &        -0.033      &    0.015\\
 & 8248.49174 &  0.127         &     0.019                 &        -0.030     &    0.014\\
 & 8249.43605 &  0.132         &     0.019                 &        -0.052      &   0.014\\
 & 8249.48646 &  0.109         &     0.019                 &        -0.022      &   0.017\\
\hline
GIANO-B & 8266.37480 &  0.610         &     0.017     &        -0.046      &         0.018 \\
(G2)  & 8266.39281 &  0.615       &     0.017  &        -0.052      &         0.016 \\
 & 8266.45613 &  0.592        &     0.017                 &        -0.025     &   0.016 \\
 & 8267.37837 &  0.586         &     0.017                 &        -0.060     &    0.015\\
 & 8267.39186 &  0.566         &     0.017                 &        -0.034     &     0.013 \\
 & 8267.40012 &  0.592       &     0.017                   &        -0.033      &     0.014 \\
 & 8268.38613 &  0.587       &     0.017                   &       -0.027      &    0.021\\
 & 8268.45429 &  0.566       &     0.017                   &       -0.027     &    0.031\\
 & 8269.39477 &  0.569       &     0.017                   &       -0.034     &  0.013\\
 & 8270.39491 &  0.583       &     0.017                   &       -0.039      &    0.013\\
 & 8275.39365 &  0.621       &     0.017                   &       -0.068      &    0.013\\
 & 8277.39237 &  0.618       &     0.017                   &       -0.065      &   0.013\\
 & 8279.43277 &  0.596       &     0.017                   &       0.024      &    0.013\\
\hline
\noalign{\smallskip}
\end{longtable}

\begin{longtable}{lcc}
\caption{\label{tab:YO17stellaV} Time series of AD Leo from STELLA data in $V$-band. For each observation we list the photometric normalized flux and its uncertainty.}\\
\hline
\noalign{\smallskip}
 JD-2450000  &      Normalized Flux         &    $\sigma_{\rm Flux}$         \\
\hline             
\noalign{\smallskip}
8225.49705 & 0.9982 & 0.0019 \\
8225.49879 & 0.9972 & 0.0019 \\
8225.50054 & 0.9987 & 0.0019 \\
8225.50229 & 0.9996 & 0.0019 \\
8235.35905 & 0.9944 & 0.0021 \\
8235.36083 & 0.9939 & 0.0021 \\
8235.36261 & 0.9905 & 0.0021 \\
8235.36439 & 0.9881 & 0.0021 \\
8238.36017 & 1.0116 & 0.0021 \\
8238.36195 & 1.0149 & 0.0020 \\
8238.36373 & 1.0131 & 0.0019 \\
8238.36550 & 1.0103 & 0.0019 \\
8239.36147 & 0.9908 & 0.0020 \\
8239.36324 & 0.9853 & 0.0019 \\
8239.36503 & 0.9933 & 0.0019 \\
8239.36681 & 0.9905 & 0.0020 \\
8242.36014 & 1.0040 & 0.0019 \\
8242.36192 & 1.0009 & 0.0019 \\
8242.36370 & 1.0056 & 0.0020 \\
8242.36547 & 1.0031 & 0.0019 \\
8243.36117 & 0.9993 & 0.0019 \\
8243.36295 & 1.0012 & 0.0020 \\
8243.36472 & 1.0006 & 0.0019 \\
8243.36650 & 0.9992 & 0.0019 \\
8246.37223 & 0.9931 & 0.0020 \\
8246.37401 & 0.9863 & 0.0019 \\
8246.37578 & 0.9877 & 0.0019 \\
8246.37756 & 0.9894 & 0.0019 \\
8247.36200 & 1.0034 & 0.0019 \\
8247.36377 & 0.9958 & 0.0019 \\
8247.36555 & 1.0035 & 0.0019 \\
8247.36732 & 1.0030 & 0.0019 \\
8248.36254 & 0.9940 & 0.0019 \\
8248.36437 & 0.9959 & 0.0019 \\
8248.36615 & 0.9924 & 0.0019 \\
8248.36792 & 0.9902 & 0.0019 \\
8249.36275 & 1.0079 & 0.0020 \\
8249.36451 & 1.0078 & 0.0019 \\
8249.36626 & 1.0043 & 0.0019 \\
8249.36802 & 1.0039 & 0.0019 \\
8250.36504 & 0.9859 & 0.0019 \\
8250.36680 & 0.9836 & 0.0019 \\
8250.36856 & 0.9807 & 0.0020 \\
8250.37031 & 0.9898 & 0.0019 \\
8253.37865 & 0.9982 & 0.0023 \\
8253.38040 & 0.9951 & 0.0022 \\
8253.38222 & 0.9920 & 0.0021 \\
8253.38398 & 0.9962 & 0.0021 \\
8257.36574 & 0.9898 & 0.0020 \\
8257.36751 & 0.9888 & 0.0020 \\
8257.36929 & 0.9899 & 0.0020 \\
8257.37107 & 0.9882 & 0.0019 \\
8258.36617 & 1.0075 & 0.0020 \\
8258.36800 & 1.0067 & 0.0020 \\
8258.36978 & 1.0034 & 0.0019 \\
8258.37156 & 1.0062 & 0.0019 \\
8262.36698 & 0.9985 & 0.0020 \\
8262.36873 & 0.9971 & 0.0020 \\
8262.37048 & 0.9983 & 0.0020 \\
8262.37223 & 1.0011 & 0.0020 \\
8263.38697 & 0.9950 & 0.0020 \\
8263.38871 & 1.0008 & 0.0020 \\
8263.39046 & 1.0045 & 0.0020 \\
8263.39221 & 1.0038 & 0.0020 \\
8266.41064 & 0.9943 & 0.0021 \\
8266.41243 & 0.9886 & 0.0021 \\
8266.41418 & 0.9921 & 0.0021 \\
8266.41592 & 0.9879 & 0.0021 \\
8267.38180 & 1.0055 & 0.0020 \\
8267.38354 & 1.0092 & 0.0020 \\
8267.38529 & 1.0048 & 0.0020 \\
8267.38704 & 1.0085 & 0.0020 \\
8269.38191 & 1.0112 & 0.0020 \\
8269.38368 & 1.0065 & 0.0020 \\
8269.38546 & 1.0089 & 0.0020 \\
8269.38723 & 1.0079 & 0.0020 \\
8270.37951 & 0.9976 & 0.0020 \\
8270.38129 & 0.9946 & 0.0020 \\
8270.38307 & 0.9957 & 0.0020 \\
8270.38484 & 0.9927 & 0.0020 \\
8271.40903 & 1.0036 & 0.0022 \\
8271.41078 & 1.0102 & 0.0022 \\
8271.41253 & 1.0022 & 0.0021 \\
8271.41428 & 1.0042 & 0.0021 \\
8272.37962 & 0.9991 & 0.0020 \\
8272.38137 & 0.9974 & 0.0020 \\
8272.38312 & 0.9997 & 0.0020 \\
8272.38487 & 0.9986 & 0.0020 \\
8273.41196 & 0.9995 & 0.0022 \\
8273.41371 & 1.0045 & 0.0022 \\
8273.41545 & 1.0087 & 0.0023 \\
8273.41720 & 0.9986 & 0.0022 \\
8274.43947 & 1.0225 & 0.0025 \\
8274.44124 & 1.0061 & 0.0024 \\
8274.44308 & 1.0165 & 0.0025 \\
8274.44485 & 1.0240 & 0.0027 \\
8276.37303 & 1.0086 & 0.0022 \\
8276.37480 & 1.0115 & 0.0022 \\
8276.37658 & 1.0083 & 0.0022 \\
8276.37836 & 1.0105 & 0.0022 \\
8277.39316 & 1.0023 & 0.0022 \\
8277.39494 & 0.9982 & 0.0022 \\
8277.39677 & 0.9994 & 0.0022 \\
8277.39855 & 0.9991 & 0.0021 \\
8279.37414 & 0.9927 & 0.0021 \\
8279.37592 & 1.0006 & 0.0021 \\
8279.37770 & 0.9962 & 0.0022 \\
8279.37948 & 0.9922 & 0.0021 \\
8281.41871 & 1.0008 & 0.0021 \\
8281.42049 & 1.0077 & 0.0021 \\
8281.42223 & 1.0038 & 0.0021 \\
8281.42399 & 0.9897 & 0.0021 \\
8283.37423 & 1.0056 & 0.0022 \\
8283.37599 & 1.0041 & 0.0021 \\
8283.37774 & 1.0148 & 0.0021 \\
8283.37950 & 1.0123 & 0.0021 \\
\hline
\noalign{\smallskip}
\end{longtable}

\begin{longtable}{lcc}
\caption{\label{tab:YO17stellaI} Time series of AD Leo from STELLA data in $I$-band. For each observation we list the photometric normalized flux and its uncertainty.}\\
\hline
\noalign{\smallskip}
 JD-2450000  &      Normalized Flux         &    $\sigma_{\rm Flux}$         \\
\hline             
\noalign{\smallskip}
8225.49792 & 0.9975 & 0.0018 \\
8225.49967 & 1.0002 & 0.0018 \\
8225.50142 & 0.9997 & 0.0018 \\
8225.50316 & 0.9971 & 0.0019 \\
8235.35994 & 1.0018 & 0.0020 \\
8235.36172 & 1.0012 & 0.0020 \\
8235.36350 & 0.9996 & 0.0020 \\
8235.36528 & 1.0005 & 0.0020 \\
8238.36106 & 1.0099 & 0.0020 \\
8238.36284 & 1.0050 & 0.0019 \\
8238.36461 & 1.0065 & 0.0019 \\
8238.36639 & 1.0041 & 0.0019 \\
8239.36236 & 0.9941 & 0.0019 \\
8239.36413 & 0.9914 & 0.0019 \\
8239.36592 & 0.9907 & 0.0021 \\
8239.36770 & 0.9937 & 0.0018 \\
8242.36103 & 1.0033 & 0.0018 \\
8242.36281 & 1.0073 & 0.0019 \\
8242.36459 & 1.0044 & 0.0018 \\
8242.36636 & 1.0016 & 0.0018 \\
8243.36206 & 1.0014 & 0.0019 \\
8243.36383 & 1.0036 & 0.0019 \\
8243.36561 & 0.9996 & 0.0018 \\
8243.36739 & 0.9967 & 0.0018 \\
8246.37312 & 0.9906 & 0.0018 \\
8246.37490 & 0.9955 & 0.0018 \\
8246.37667 & 0.9929 & 0.0018 \\
8246.37845 & 0.9906 & 0.0019 \\
8247.36288 & 1.0008 & 0.0018 \\
8247.36466 & 0.9999 & 0.0018 \\
8247.36644 & 1.0070 & 0.0018 \\
8247.36822 & 1.0035 & 0.0018 \\
8248.36343 & 0.9990 & 0.0018 \\
8248.36526 & 0.9938 & 0.0018 \\
8248.36703 & 0.9944 & 0.0018 \\
8248.36881 & 0.9956 & 0.0018 \\
8249.36363 & 0.9978 & 0.0019 \\
8249.36538 & 1.0032 & 0.0019 \\
8249.36714 & 0.9998 & 0.0019 \\
8249.36890 & 0.9991 & 0.0019 \\
8250.36592 & 1.0018 & 0.0019 \\
8250.36768 & 0.9945 & 0.0019 \\
8250.36943 & 0.9936 & 0.0019 \\
8250.37119 & 0.9955 & 0.0019 \\
8253.37953 & 0.9889 & 0.0022 \\
8253.38134 & 0.9974 & 0.0019 \\
8253.38310 & 0.9941 & 0.0019 \\
8253.38485 & 0.9980 & 0.0020 \\
8257.36662 & 0.9977 & 0.0019 \\
8257.36840 & 0.9966 & 0.0019 \\
8257.37018 & 1.0028 & 0.0019 \\
8257.37196 & 0.9963 & 0.0019 \\
8258.36712 & 1.0003 & 0.0019 \\
8258.36889 & 1.0003 & 0.0019 \\
8258.37067 & 1.0030 & 0.0019 \\
8258.37245 & 1.0085 & 0.0019 \\
8262.36785 & 0.9953 & 0.0020 \\
8262.36960 & 0.9971 & 0.0020 \\
8262.37135 & 0.9992 & 0.0020 \\
8262.37310 & 1.0074 & 0.0019 \\
8263.38784 & 1.0017 & 0.0019 \\
8263.38959 & 1.0013 & 0.0019 \\
8263.39133 & 0.9959 & 0.0019 \\
8263.39308 & 1.0003 & 0.0019 \\
8266.41156 & 0.9945 & 0.0020 \\
8266.41331 & 0.9938 & 0.0020 \\
8266.41505 & 1.0014 & 0.0020 \\
8266.41679 & 0.9915 & 0.0020 \\
8267.38267 & 1.0052 & 0.0019 \\
8267.38442 & 1.0114 & 0.0019 \\
8267.38616 & 1.0091 & 0.0019 \\
8267.38791 & 0.9999 & 0.0019 \\
8269.38279 & 1.0011 & 0.0019 \\
8269.38457 & 1.0019 & 0.0019 \\
8269.38635 & 0.9996 & 0.0019 \\
8269.38812 & 0.9960 & 0.0019 \\
8270.38040 & 0.9966 & 0.0019 \\
8270.38218 & 1.0007 & 0.0020 \\
8270.38396 & 0.9979 & 0.0019 \\
8270.38573 & 0.9908 & 0.0019 \\
8271.40990 & 1.0044 & 0.0020 \\
8271.41165 & 1.0157 & 0.0020 \\
8271.41340 & 0.9952 & 0.0020 \\
8271.41515 & 0.9963 & 0.0020 \\
8272.38050 & 1.0030 & 0.0019 \\
8272.38225 & 1.0117 & 0.0019 \\
8272.38400 & 1.0131 & 0.0018 \\
8272.38575 & 0.9959 & 0.0019 \\
8273.41283 & 1.0037 & 0.0021 \\
8273.41458 & 0.9965 & 0.0021 \\
8273.41633 & 1.0042 & 0.0021 \\
8273.41807 & 1.0008 & 0.0021 \\
8274.44036 & 1.0049 & 0.0023 \\
8274.44219 & 1.0012 & 0.0023 \\
8274.44396 & 1.0239 & 0.0024 \\
8274.44574 & 1.0051 & 0.0025 \\
8276.37392 & 0.9977 & 0.0021 \\
8276.37569 & 1.0047 & 0.0021 \\
8276.37747 & 1.0061 & 0.0021 \\
8276.37925 & 1.0099 & 0.0020 \\
8277.39405 & 0.9955 & 0.0020 \\
8277.39583 & 0.9933 & 0.0020 \\
8277.39766 & 0.9946 & 0.0020 \\
8277.39944 & 0.9983 & 0.0020 \\
8279.37503 & 1.0008 & 0.0020 \\
8279.37681 & 1.0087 & 0.0020 \\
8279.37859 & 0.9994 & 0.0020 \\
8279.38036 & 1.0023 & 0.0020 \\
8281.41959 & 0.9925 & 0.0019 \\
8281.42136 & 0.9932 & 0.0020 \\
8281.42311 & 0.9926 & 0.0020 \\
8281.42486 & 0.9973 & 0.0020 \\
8283.37511 & 1.0001 & 0.0020 \\
8283.37686 & 0.9986 & 0.0020 \\
8283.37861 & 1.0066 & 0.0020 \\
8283.38037 & 1.0011 & 0.0019 \\
\hline
\noalign{\smallskip}
\end{longtable}

\end{appendix}
\twocolumn

\end{document}